\documentclass[sn-mathphys,iicol]{sn-jnl}
\usepackage[noadjust]{cite}

\DeclareUnicodeCharacter{2009}{FIXME}

\jyear{2022}%

\theoremstyle{thmstyleone}%
\theoremstyle{thmstyletwo}%

\theoremstyle{thmstylethree}%

\begin{document}

\title[Article Title]{High-energy betatron source driven by a 4-PW laser with applications to non-destructive imaging}


\author*[1]{\fnm{Calin Ioan} \sur{Hojbota}}
\email{calinh@ibs.re.kr}

\author[1]{\fnm{Mohammad} \sur{Mirzaie}}
\author[1]{\fnm{Do Yeon} \sur{Kim}}
\author[1,2]{\fnm{Tae Gyu} \sur{Pak}}
\author[1,3]{\fnm{Mohammad} \sur{Rezaei-Pandari}}
\author[1,4]{\fnm{Vishwa Bandhu} \sur{Pathak}}
\author[1]{\fnm{Jong Ho} \sur{Jeon}}
\author[1,5]{\fnm{Jin Woo} \sur{Yoon}}
\author[1,5]{\fnm{Jae Hae} \sur{Sung}}
\author[1,5]{\fnm{Seong Ku} \sur{Lee}}
\author[1,5]{\fnm{Chul Min} \sur{Kim}}
\author[1,2,6]{\fnm{Ki Yong } \sur{Kim}}
\author[1,2]{\fnm{Chang Hee} \sur{Nam}}



\affil[1]{\orgdiv{Center for Relativistic Laser Science},  \orgname{Institute for Basic Science}, 
			\orgaddress{\city{Gwangju}, \postcode{61005}, \country{Republic of Korea}}}

\affil[2]{\orgdiv{Dept. of Physics and Photon Science},  \orgname{Gwangju Institute of Science and Technology}, 
			\orgaddress{ \city{Gwangju}, \postcode{61005}, \country{Republic of Korea}}}
			
\affil[3]{\orgdiv{Laser and Plasma Research Institute},  \orgname{Shahdi Beheshti University}, 
			\orgaddress{\city{Teheran},  \country{Iran}}}
			
\affil[4]{\orgdiv{School of Advanced Science},  \orgname{Vellore Institute of Technology}, 
			\orgaddress{\city{Vellore}, \postcode{632014}, \state{Tamilnandu}, \country{India}}}
			
\affil[5]{\orgdiv{Advanced Photonics Research Institute},  \orgname{Gwangju Institute of Science and Technology}, 
			\orgaddress{\city{Gwangju}, \postcode{61005}, \country{Republic of Korea}}}
			
\affil[6]{\orgdiv{Institute for Research in Electronics and Applied Physics},  \orgname{University of Maryland}, 
			\orgaddress{ \city{College Park}, \postcode{20742}, \state{Maryland},\country{USA}}}



 
\abstract{Petawatt-class lasers can produce multi-GeV electron beams through laser wakefield electron acceleration. As a by-product, the accelerated electron beams can generate broad synchrotron-like radiation known as betatron radiation. In the present work, we measure the properties of the radiation produced from $2$ GeV, $215$ pC electron beams, which shows a broad radiation spectrum with a critical energy of $515$ keV, extending up to MeV photon energies and $10$ mrad divergence. Due to its high energy and flux, such radiation is an ideal candidate for $\gamma$-ray radiography of dense objects. We employed a compact betatron radiation setup operated at relatively high-repetition rates ($0.1$ Hz) and used it to scan cm-sized objects: a DRAM circuit, BNC and SMA connectors, a padlock and a gas jet nozzle. GEANT4 simulations were carried out to reproduce the radiograph of the gas jet. The setup and the radiation source can reveal the interior structure of the objects at the sub-mm level, proving that it can further be applied to diagnose cracks or holes in various components. The radiation source presented here is a valuable tool for non-destructive inspection and for applications in high-energy-density physics such as nuclear fusion.}




\maketitle

\section{Introduction}
\label{sec1}

Due to the recent proliferation of petawatt (PW) lasers \citep{bib1}, laser wakefield electron acceleration (LWFA) has rapidly established itself as the most promising technique for building compact particle accelerators. During the propagation of an intense (intensity $>10^{18}$ W/cm$^2$) laser pulse inside an underdense plasma medium, a space-charge cavity is formed behind the pulse, exhibiting longitudinal fields of $100$'s GV/m. Such large fields are orders of magnitude higher than the ones produced in RF LINACS, which allows for the acceleration of electrons to multi-GeV energies within centimeters \citep{bib2,bib3}. So far, progress in this area has led to the production of electron beams having energies up to $8$ GeV \citep{bib4}, and more recently $>10$ GeV \citep{bib5}. The beams can have up to micro-coulomb charge \citep{bib6}, can achieve $>1\%$ energy spreads \citep{bib7}, and can even be produced at kHz repetition rates \citep{bib8,bib9}. Such properties make these accelerators suitable drivers for cutting-edge radiation sources \citep{bib10}, as recently demonstrated in LWFA-based free-electron lasers \citep{bib11} and Thomson/Compton scattering sources \citep{bib12,bib13}.

During the acceleration process, the electron beams with a large Lorentz factor ($\gamma>1000$) can undergo transverse oscillations \citep{bib14} due to the transverse focusing force of the space-charge cavity on the electrons. This process leads to the generation of a broad synchrotron-like x-ray spectrum \citep{bib15}, known in the community as betatron radiation \citep{bib15,bib16}. This radiation source is a very attractive alternative to large synchrotron machines, as it can produce a large photon number ($\gt 10^8$); it is collimated (typically up to a few 10s of mrad), and its energy varies from a few-keV up to hundreds of keV.  In addition, these sources can be studied in small-scale laser setups and at PW facilities, making them a cheaper alternative to synchrotrons. One unique feature of such a source is its small emission size ($\approx 1$ $\mu$m), which allows micron-scale imaging \citep{bib17}. So far, betatron sources have been used for phase-contrast imaging \citep{bib18,bib19}, tomography \citep{bib20,bib21}, diagnostics of shocks produced in laser-plasma \citep{bib22} and have been proposed for the diagnostic of fusion plasmas \citep{bib23}. Despite many potential applications, most of the radiographs are produced with energies $\lt 100$ keV (for example \citep{bib24}), while high-energy applications in the hard x-ray spectrum ($0.1-1$ MeV) are rather rare due to the sparsity of high energy ($\gt 1$ GeV) LWFA beams required to produce such a spectrum. 

One potential application of high-energy betatron sources is non-invasive radiographic diagnostics of high-density objects, an area sometimes referred to as non-destructive imaging. For such applications, $\gamma$-ray sources are of great interest as they can probe the interior structure of objects without disassembling the devices or cutting through them. For comparison, we note that bremsstrahlung radiation has already been employed \citep{bib25}, yet such sources require additional steps to produce $\gamma$-ray radiation. On the other hand, betatron radiation generation is an inherent process during LWFA.  However, LWFA-based betatron radiation has yet to demonstrate industrial $\gamma$-ray radiography. This was highlighted as one of the future research directions on LWFA-based radiation sources, and this aspect is of significant importance for the community \citep{bib26}. 

The present work, therefore, aims to produce and apply high-energy betatron radiation generated with the CoReLS 4-PW laser to the radiographic investigation of industrial objects. In the following sections, we present the setup used for LWFA and betatron generation and demonstrate a detection setup to diagnose the radiation produced in such a work. We measure the beam properties, such as spectrum and divergence, and deduce other parameters of the radiation. We then demonstrate successful radiography of several objects, including the realization of single-shot radiography, with $\lt 200$ $\mu$m spatial resolution. Finally, we show how such a source can be used to extract information about the interior size of the objects and correlate our measurements with GEANT4 simulations. This work is a further step in the production of high-energy betatron radiation and its application for the radiographic investigation of dense objects and non-destructive imaging.


\section{Measurement of betatron radiation produced with the 4-PW laser}\label{sec2}

 \subsection{Experimental setup}\label{subsec2}

\begin{figure*}[h]%
\centering
\includegraphics[width=0.99\textwidth]{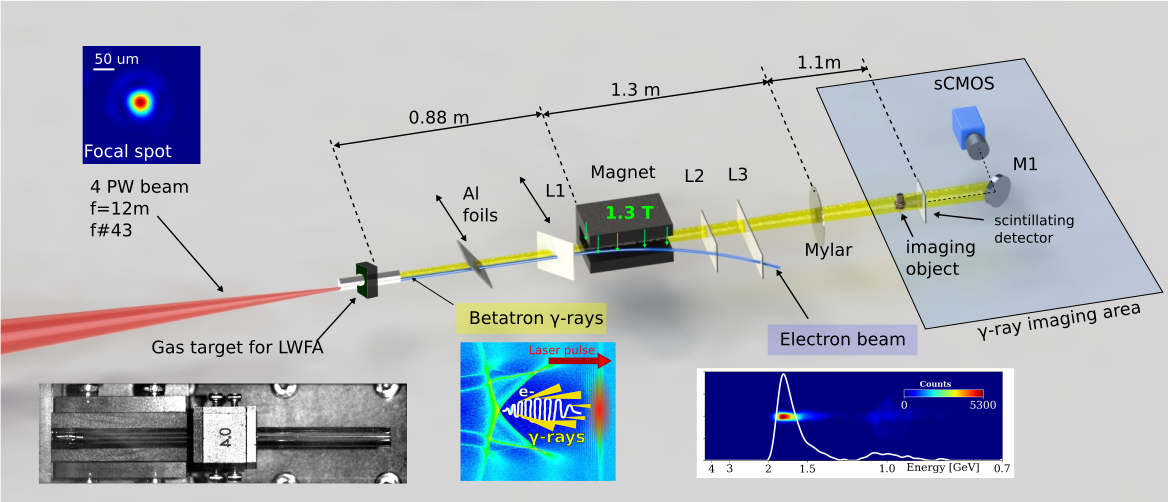}
\caption{
LWFA and radiography setup. The 4-PW beam (red) enters from the left side and is focused down to 45 $\mu$m (Focal spot, inset), hitting the gas target for LWFA. High energy electrons oscillate, producing betatron $\gamma$-rays (yellow beam). The L2 and L3 diagnose the energy of the electron beam (shown in the inset), while the radiation beam exits the vacuum chamber through the Mylar window. The radiography is performed using the scintillating detector and the sCMOS camera in the $\gamma$-ray imaging area.  
}
\label{fig1}
\end{figure*}

The experiments presented in this work were carried out using the 4-PW laser at CoReLS \citep{bib27}. The system can produce $20$ fs laser pulses with up to $80$ J energy and percent-level energy stability. After exiting the compression stage, the laser beam was transported to the experimental area, where it was focused by a concave mirror (f=$12$ m and f/$43$) to a spot of 49 $\mu$m, producing on-target an intensity of $1.9\times 10^{19}$ W/cm$^2$. The intensity corresponded to a normalized vector potential , where  is the electron charge,  is the peak vector potential of the laser,  is the rest mass of the electron, and  represents the speed of light. The pulse was focused inside a 10 cm long gas target, filled with He (97$\%$) and Ne (3$\%$). This particular gas combination was found to be suitable for ionization injection \citep{bib28} in the self-truncated regime \citep{bib29}. In addition, the laser chirp was controlled to optimize the electron beam quality \citep{bib30}. The experimental setup included an aluminum (Al) screen (consisting of foils with a combined thickness of 300 $\mu$m) to reflect the residual laser light to a beam dump and a setup consisting of a scintillating Lanex screen to monitor the electron beam pointing and divergence (L1), a magnet ($1.3$ T) that disperses the electron beam, and two subsequent Lanex screens (L2 and L3) for energy measurements. This multi-screen setup is important for calibrating electron beams that exhibit significant off-axis pointing \citep{bib31,bib32}. The peak electron density of the gas medium was estimated to be around $1.5-2.5 \times 10^{18}$ cm$^{-3}$, which is higher than the typical operating density for this accelerator ($5-8\times 10^{17}$ cm$^{-3}$). This higher density allowed the generation of a strong betatron flux. Using this setup, we produced electron beams of up to $2$ GeV, such as the one shown in Fig. 1 which is quasi-monoenergetic, having an energy spread of $\delta E/E \approx 6.3\%$ and a divergence of $2.3$ mrad (here $\delta E$ represents the standard deviation of the energy spread). The average charge of shots used in betatron experiments was $215\pm 15$ pC. The betatron beams produced by these electrons were further characterized and used for imaging.

The present imaging system was designed for the efficient acquisition of $\gamma$-ray radiographs. During the measurements of betatron radiation, the Al foils and Lanex L1 were removed in order to avoid the generation of bremsstrahlung by the electron beam. The $\gamma$-ray beam exited the target chamber through a thin $300$ $\mu$m Mylar window (see Fig. 1), which provides negligible attenuation. The deflected electron beam was dumped outside the chamber, and the imaging setup was shielded with lead bricks (not shown in Fig. 1) to block background radiation. For the imaging, we used a scintillating detector (Hamamatsu GPXS J13113) along with a 16-bit sCMOS (PCO edge). The detector screen consisted of a $400-\mu$m thick CsI ($45\times 45$ mm$^{2}$), layered over an Al substrate (500 $\mu m$). This setup was placed $3.3$ m away from the source in order to allow the diverging beam to image cm-sized objects. The objects were placed $5$ cm in front of the detector, providing almost unit magnification (M $\approx 1.02$) from the x-ray source. The purpose was not to magnify the objects due to the limited size of the detector. This setup configuration is useful for betatron imaging because it can provide a much faster response time than imaging plates (IPs): while IPs require reading over $\approx 0.5$ hour, this setup can work at high repetition rates (up to kHz). Additionally, it can cover a wide dynamic range ($\gt10^3$), providing a good contrast if required. The spatial resolution can be as low as a few 10s of microns, comparable to that of IPs \citep{bib33}. This setup allowed us to characterize betatron beams and use them for radiography rapidly.

\subsection{Betatron beam properties}\label{subsec2}
The setup shown in Fig. 1 was used to produce betatron radiation during LWFA and to characterize the spectral distribution, critical energy, divergence, and photon number. The betatron radiation spectrum is characterized by a frequency ($\omega$) distribution of the intensity ($I$) over the solid angle ($\Omega$):

\begin{align}
\frac{d^2I}{d\omega d\Omega} &\approx N_\beta \frac{3e^2}{2\pi^3 \epsilon_0 c} \frac{\gamma^2 \zeta^2 }{1+\gamma^2\theta^2} \times \nonumber
	\\ &\times \left[ \frac{\gamma^2 \zeta^2 }{1+\gamma^2\theta^2} \cdot K^2_{1/3}\left(\zeta\right) + K^2_{2/3}\left(\zeta\right) \right] \label{eq1}
\end{align}

 where  $N_\beta$ represents the number of oscillations per electron in the accelerating medium and $\epsilon_0$ is the vacuum permittivity, while $K_{1/3}$ and $K_{1/3}$ are the modified Bessel functions of the second kind. The argument $\zeta$ of the Bessel function is given by $\zeta=\omega/\omega_c (1+\gamma^2\theta^2)$ for an observation angle $\theta$ and a critical frequency $\omega_c$, which is the most important parameter as it defines the energy range of the spectrum. The corresponding critical energy, $E_c$ , can be approximated as
$E_c [keV] = 1.1 \times 10^{-23}\gamma^2 n_e [cm^{-3}]r_\beta[\mu m]$. Here, $n_e$ is the plasma density, and $r_\beta$ is the transverse oscillation radius of the electron emitting betatron radiation.  The on-axis radiation spectrum can be approximated as $\frac{dI^2}{d\omega d\Omega} \approx N_\beta \frac{3e^2}{2\pi^3\epsilon_0 c} \gamma^2\zeta^2 K^2_{2/3}\left(\zeta\right)$ while $\zeta$ reduces to $\omega/\omega_c $ (or $E/E_c$). The number distribution is straightforwardly obtained as $\frac{dN}{dE}=\frac{1}{E}\frac{dI}{dE}\zeta K^2_{2/3}(\zeta)$, which is defined by its critical energy. We can thus use a functional form of the spectrum, fully defined by a critical energy ($E_c$).

\begin{figure*}[h]%
\centering
\includegraphics[width=0.99\textwidth]{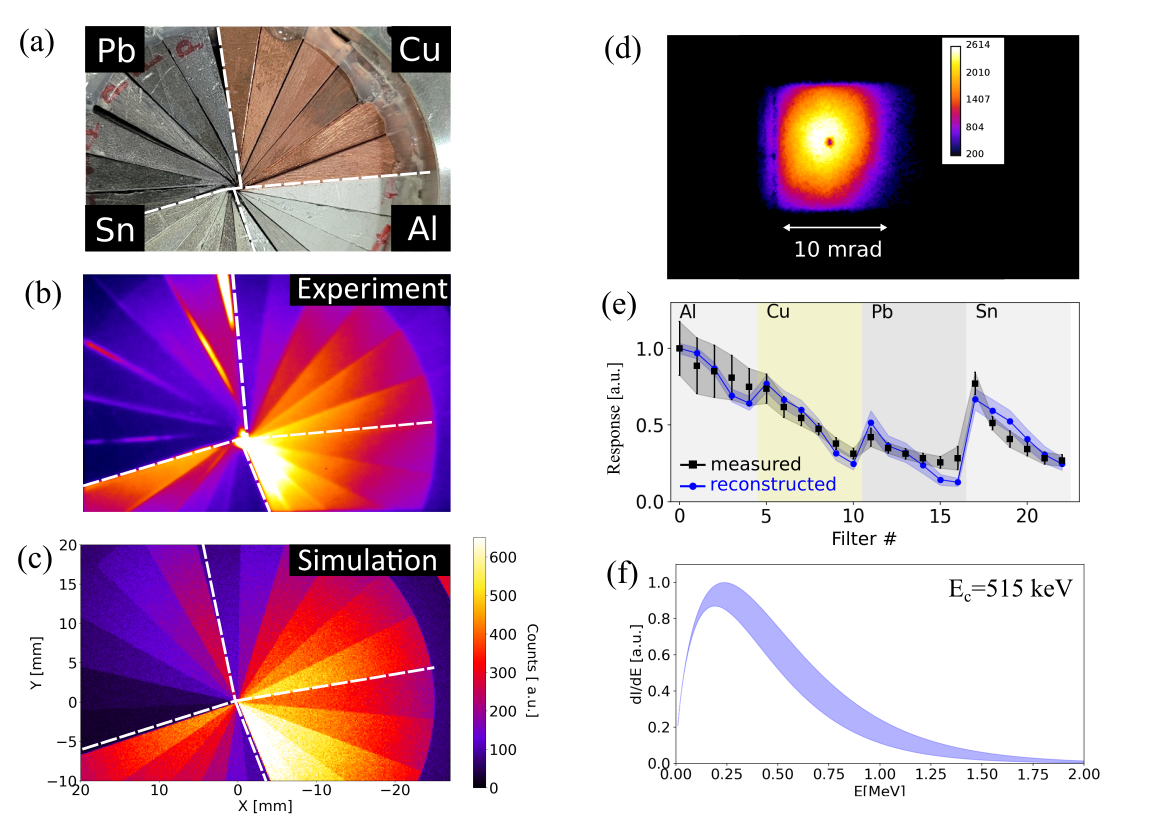}
\caption{
Diagnostic of betatron radiation. (a) Photo of the range filter stack; (b) the average of the filter response accumulated for 10 shots; (c) a GEANT4 simulation of the filter response for $E_c=500$ keV; (d) betatron beam divergence captured on L3; (e) measured and reconstructed filter response for each slice; (f) the corresponding intensity spectrum with an average critical energy $E_c=515$ keV.  
}
\label{fig2}
\end{figure*}
 The spectrum was extracted using the range filter (RF) shown in Fig. 2(a). The RF wheel consists of Al, Cu, Sn, and Pb slices, with thicknesses varying from 1 mm to 2 cm, which can differentially attenuate the spectrum. More details of this filter can be found in \citep{bib34}. Note that similar techniques are becoming a standard practice for the measurement of keV-range radiation in laser-plasma experiments \citep{bib35,bib36}. When the filter is coupled to the present setup, it can work as a faster and more convenient alternative to stack detectors that rely on IPs \citep{bib37,bib38}. For the present work, the angular dependence of the spectrum was not characterized, although such effects were likely present \citep{bib39}. We extracted the spectrum by minimizing the difference between the detector response and the calculated one in GEANT4 \citep{bib40}, adopting a functional form $\frac{dI}{dE} ~ \zeta^2 K^2_{2/3}\left(\zeta\right) $.  

	In Fig. 2(b), we present an example of a measured signal (averaged over 10 betatron shots) whose response was used to reconstruct the radiation spectrum. We obtained a critical energy of $515\pm 55$ keV after the extraction procedure. The spectrum is shown in Fig. 2(f): the band represents an uncertainty corresponding to $\pm 55$ keV in the critical energy, and the intensity range depends on the uncertainty in the $\gamma$-ray photon number. The measured and reconstructed filter responses in Fig. 2(e) show reasonable agreement with a fitting error within $\lt 10\%$. The bands represent fluctuations (one standard deviation) around the average value of the filter signal at each slice. The slight disagreement occurred due to the imperfect signal shape on the detector, asymmetric emission of the radiation, off-axis effects, and limitations on the detector size. The filter response obtained from a GEANT4 simulation is shown in Fig. 2(c). The simulation was performed using a critical energy of 500 keV and a 10-mrad divergence (similar to the experimental parameters). One can observe the significant penetration through lead and the high (almost uniform) signal through the Al section of the filter. These effects are expected from high-energy photons, and such features are confirmed in both experiment and simulation. Overall, the measurement results show good agreement among the experimental result, reconstruction, and simulation of the filter response.
	
	The betatron divergence was measured with the third Lanex screen (L3), as shown in Fig. 2(d), where 7 shots with betatron signals were averaged. The divergence was found to be $\theta_\beta^{(x)} \times \theta_\beta^{(y)}=11 $mrad $\times 9.6 $mrad with an average of $(\theta_\beta^{(x)} \times \theta_\beta^{(y)})^{1/2} \approx 10 mrad$  .  For betatron beams, the emission half-angle is given as $\theta_\beta = K_\beta /\gamma $, where $K_\beta$ represents the undulator parameter of the emission and puts an upper limit on the opening angle of the radiation. For a series of shots with an average $\gamma \approx 3000$, it follows that $K_\beta  \approx 30$. When $K_\beta \gg 1$ the radiation spectrum is continuous, and when $K_\beta \ll 1$, the signal exhibits multiple harmonics. The large value of $K_\beta$, obtained from the experiment, confirmed that we can apply the continuous spectrum model described earlier. 
	
	In addition to the validity of the spectrum used in our experiments, $K_\beta$, can be used to deduce other parameters of the source. Considering that $K_\beta \approx 1.33 \times 10^{-10} \sqrt{ \gamma n_e [cm^{-3}]}r_\beta[\mu$m$]$, we can extract an oscillation radius of $r_\beta=2.8$ $\mu$m. By using the previous estimation of the critical energy and considering the experimental density range, we can also deduce $E_c =443-665$ keV, which is in good agreement with $E_c = 515$ keV extracted from the filter response. In addition, the number of emitted photons can be further estimated. The total number of $\gamma$-rays that each electron emits can be defined as $N_{\gamma/e-} = 5.6\cdot 10^{-3} N_\beta K_\beta$, where $N_\beta$ stands for the number of oscillations each electron undergoes [14]. Considering the betatron period, $ \lambda_\beta = 1.8$ mm, and a propagation range of 1.8-18 mm (i.e. 1-10 periods), we can assume $N_\beta =1.1-11$, giving a total number of $\gamma$-rays emitted by each electron,$N_\gamma = 0.2-2$ . Thus, for an electron beam of $215 \pm 15$ pC, we expect  $10^8-10^9$ $\gamma$-ray photons. For example, for two of the shots we measured, we estimated $2\cdot 10^9$ and $5 \cdot 10^9$ photons, respectively, taking into account the properties of the detection setup (geometry, QE, scintillation and absorption efficiency, etc.). The high photon number from both analytic and experimental estimations was consistent with what we observed in the imaging experiments since a much smaller photon number (e.g., $10^6$) would make it challenging to observe a reasonable image quality, especially for single-shot measurements. We note that the overall estimates agreed with the experimental observations, confirming that this source is well-suited for imaging applications.

\section{$\gamma$-ray radiography using a high-energy betatron source}\label{sec3}

\subsection{Non-destructvie imaging of high-density targets}\label{subsec3}
The betatron source and the setup in Fig. 2 were used to perform $\gamma$-ray radiography of cm-scale objects. Objects with high-density elements (up to $11 $g/cm$^3$ and effective atomic number $Z_{eff} <82$) were chosen in order to examine the appropriateness of the source and setup for industrial radiography. We placed the objects close to the detection screen to ensure a large areal density of the photons. When the beam hits the detector area, the expected areal density of photons at the object location was $1.9-5.5 \cdot 10^6 [$ph/mm$^2]$. The flux of the source was estimated to be $3.1-8.4\times 10^{10} [$ph/keV/Sr$]$at 200 keV, comparable to a bremsstrahlung source produced by multi-MeV electrons \citep{bib41}, implying that the source is suitable for imaging dense targets. A large photon flux and a large photon areal density are required for high penetration inside materials, producing a good spatial resolution and a high contrast for the radiographs. The spatial resolution was limited to $38.5$ $\mu$m/pixel, a limit restricted by the optical sensor and optical imaging system magnification. The minimum resolvable feature was $\approx 150$ $\mu$m. This configuration ensured the measurement of sub-mm features while keeping the setup large enough to distinguish various macroscopic features.

\begin{figure*}[h]%
\centering
\includegraphics[width=0.99\textwidth]{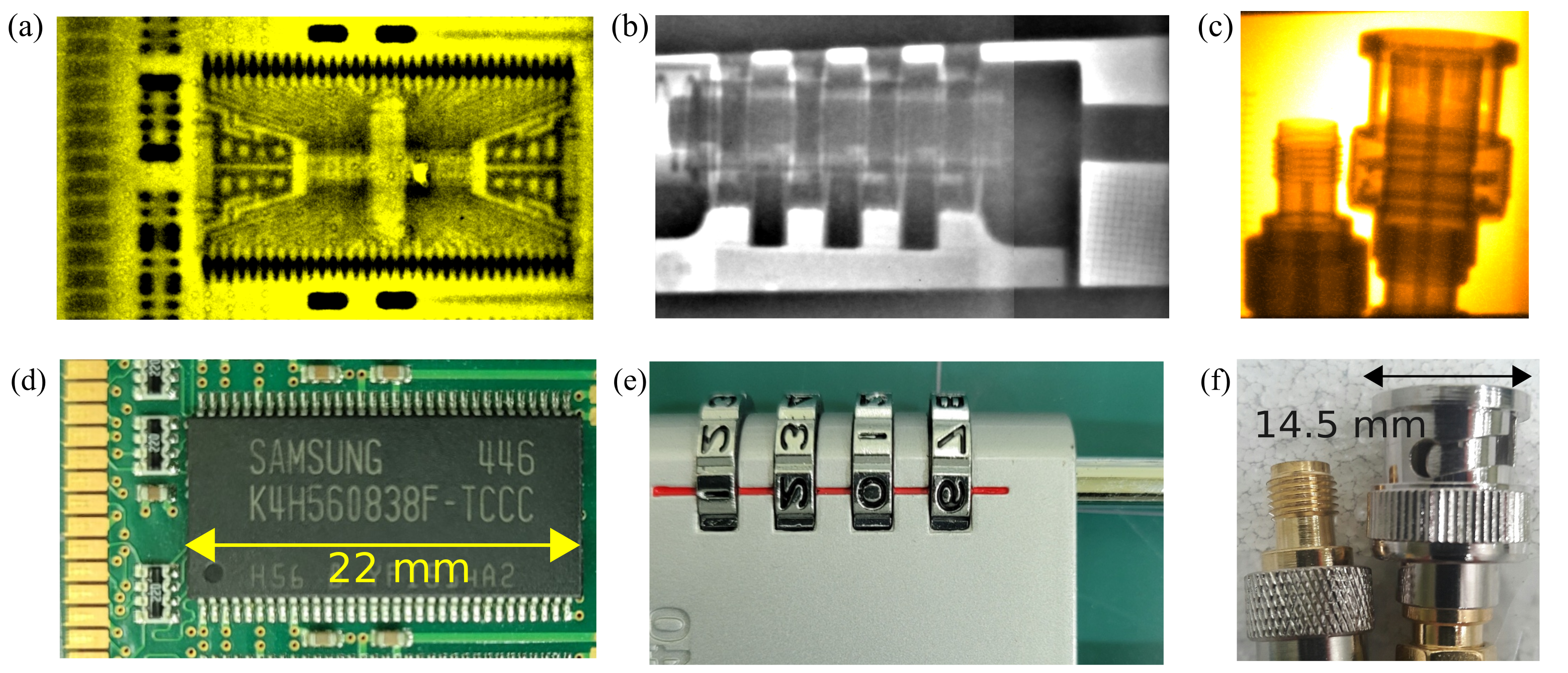}
\caption{
$\gamma$-ray imaging using betatron radiation.  Radiographs (a)-(c) are shown with the corresponding object photos (d)-(f). (a) and (d): a DRAM circuit, (b) and (e) a padlock, (c) and (f): SMA (left) and BNC (right) connectors. 
}
\label{fig3}
\end{figure*}

\begin{figure*}[h]%
\centering
\includegraphics[width=0.99\textwidth]{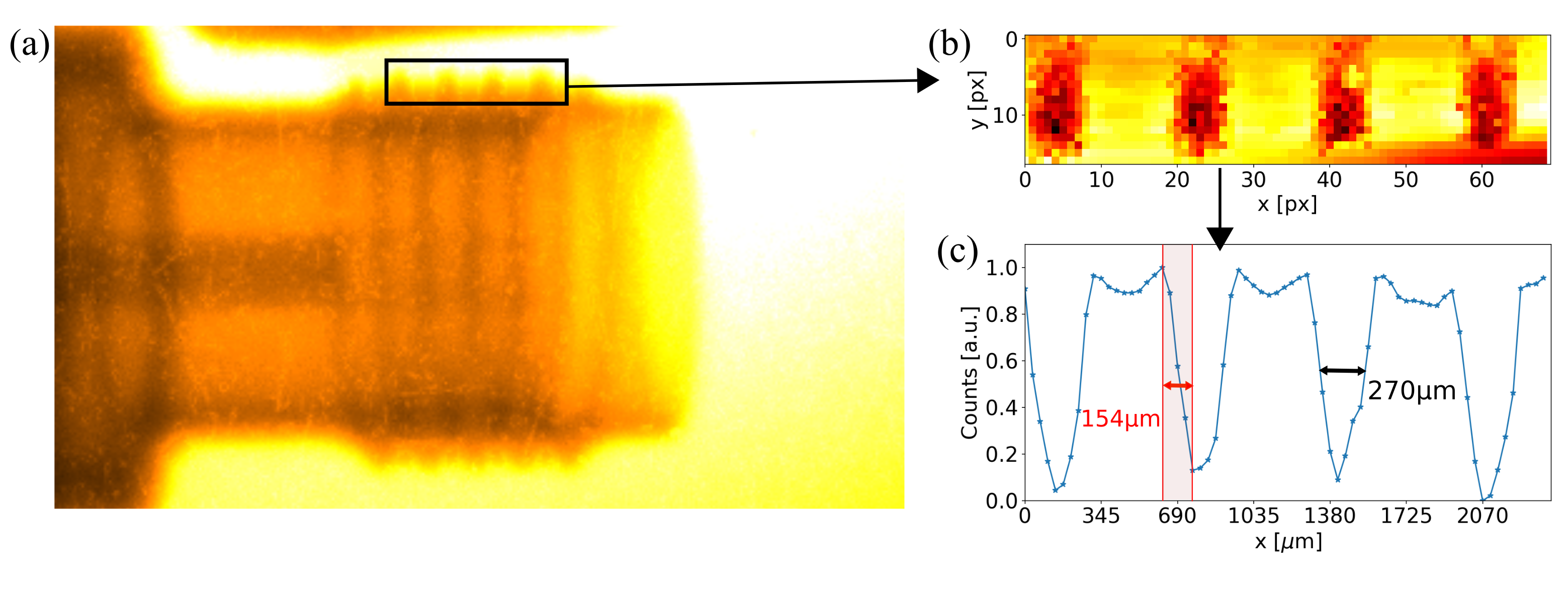}
\caption{
Spatial resolution. (a) magnified SMA connector; (b) threads of the SMA connector head; (c) lineout of the threads, showing the edge detail ($154$ $\mu$m) and the FWHM thread width ($270$ $\mu$m).}
\label{fig4}
\end{figure*}
	The objects imaged in this configuration, together with their corresponding object images, are shown in Fig. 3. The radiograph of a DDR memory chip is presented in Fig. 3(a). The radiography of such circuits is motivated by the growth in the manufacturing volume of integrated circuits, whose counterfeiting practice represents a global threat \citep{bib42}. The image was obtained by accumulating 3 shots, and Fourier filtering was used to remove large spatial components (i.e., accounting for the shape of the beam). Typically, such radiography of thin objects requires lower $\gamma$-ray energy ($40-100$ keV \citep{bib43}). For the current source with $E_c=515$ keV, the average energy would be $\approx 220$ keV; thus, this source is highly penetrating for the object tested in here, reducing the imaging contrast. It is noted that multiple acquisitions could significantly improve the imaging quality. Nevertheless, the connectors and wires inside the object are clearly visible. The interior structure of the chip may help to distinguish counterfeit components based on their geometric shape and layout \citep{bib44}.  	

	The second and third radiographic images were acquired, considering applications related to manufacturing. The second radiography shown in Fig. 3(b) corresponds to a padlock (ZARKER XD40) that has a metallic alloy casing. Through this radiography, the internal wheels, connection, the spring at the bottom, and the connecting rod are visible. The image was assembled from several successive shots taken at different positions. The image in Fig. 3(c) shows BNC (right) and SMA (left) cable connectors. Such radiography typically requires an average energy of about $100$ keV \citep{bib45, bib46}, but a higher energy would also be suitable due to a thick object size (i.e., compared to the circuit). Threads, cables, and moving parts are all visible in the images acquired in Fig. 3(c). The betatron source can thus be used to reveal the layers of assembly, internal wiring, and connected components. Such diagnostics can easily reveal defects and wrong structures and assess the precision of assembled or soldered parts during production. 
	
	An evaluation of the spatial resolution is presented in Fig. 4. For the SMA connector (Fig. 3 (c)), the threads are clearly visible, as shown in Fig. 4 (a). The corresponding lineout of the image (Fig. 4 (c)) shows that the size of a thread is around $\approx 270$ $\mu$m in FWHM (7 pixels), while the edge resolution is $\approx 4$ pixels or $\approx 150$ $\mu$m. Here the spatial resolution was mainly determined by the geometrical alignment of the object and the choice of detection setup since, otherwise, the betatron sources are known to allow for a spatial resolution below $10$ $\mu$m. We note that a very thin structure or an extremely precise alignment to the betatron axis is necessary to observe sharp edges with micron-level widths. From simple geometrical estimations, a tilt angle of $>1^{\circ}$ for an object of 5 mm thickness would decrease the sharpness of its imaging, making it impossible to observe the edge with precision higher than $ 100$ $\mu$m. From the considerations above, we can conclude that for cm-sized objects such as the ones shown in Fig. 3(b) and 3(c), the present configuration setup can achieve sub-mm resolutions of $\approx 150$ $\mu$m. 

\begin{figure*}[h]%
\centering
\includegraphics[width=0.99\textwidth]{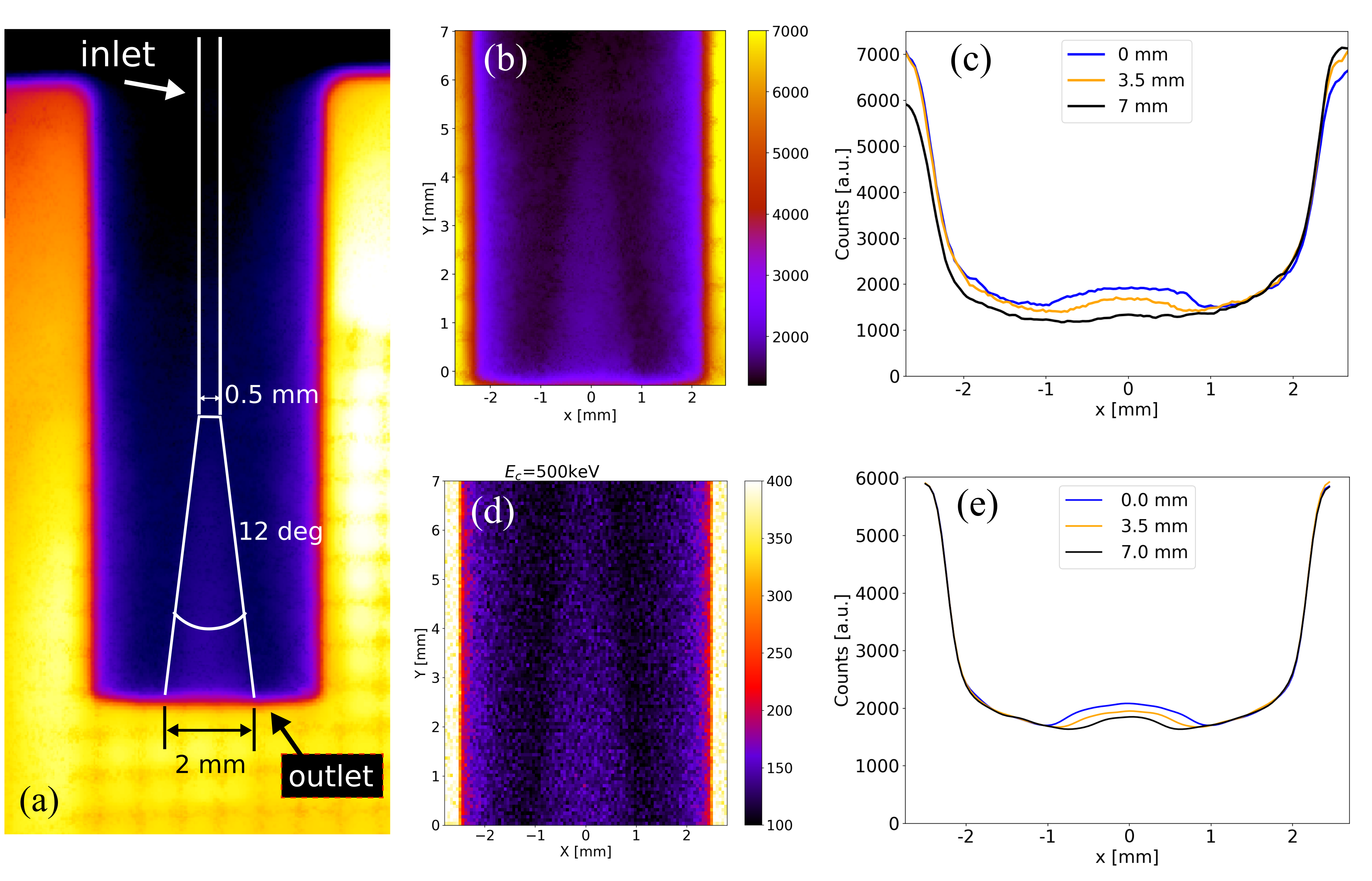}
\caption{
ingle-shot radiographic imaging of a gas jet nozzle. (a) Properties of gas nozzle: inlet size, outlet size, and opening angle obtained from the radiograph. (b) Experimental and (d) simulated radiographs with their corresponding lineouts plotted at different heights in (c) and (e).  The lineout signal in the simulation (d) was adjusted ($\times 15$) for a better comparison with the experiment. }\label{fig5}
\end{figure*}

\subsection{Single-shot extraction of gas jet properties}\label{subsec3}
	As another example, we investigated the properties of a gas jet nozzle using $\gamma$-ray radiography. The gas jet nozzle was made of brass (density $8.7 $g/cm$^3$, effective atomic number $Z_eff \approx 29-39$) and custom manufactured for LWFA experiments. The device (seen in Fig. 5(a) radiograph) has an outer diameter of $5$ mm and has an internal structure (drilled) through which the gas flows: an inner hole in which the gas enters the nozzle (inlet) and an exit orifice (outlet) through which the gas is puffed out (bottom part of the image). The outlet opens up in a conical shape, allowing the gas to flow at supersonic speeds.  Errors during the drilling process can result in undesired gas flows and gas density profiles. It is important to confirm the inner structure of a manufactured nozzle, such as the opening angle or the dimensions of the inlet and outlet. Therefore, this gas nozzle is an ideal object of investigation for demonstrating the applicability of the current betatron source.
	In Fig. 5, we analyze in detail the properties of the gas nozzle and reproduce the experimental data using GEANT4 simulations. The diameter and the opening angle of this gas jet were retrieved in a single shot, as shown in Fig. 5(a). At the exit (outlet), the orifice diameter is 2 mm with an opening angle of $12^\circ$, which we can extract from the image, while the inlet diameter narrows to $500  \mu$m. Figures 5(b)-(e) further confirm the experimental result. The simulated radiograph of the nozzle (Fig. 5 (d)) has a similar visibility compared to the experimental one (Fig. 5 (b)). For the simulation, we used a $\gamma$-photon spectrum with $E_c=500$ keV, $N_\gamma =10^7 $ particles, and a large beam size, having a photon areal density at the object plane of $10^5 ph/mm^2$. Using these parameters, we could accurately reproduce the image obtained in the experiment. 
	
	\begin{figure*}%
\centering
\includegraphics[width=0.99\textwidth]{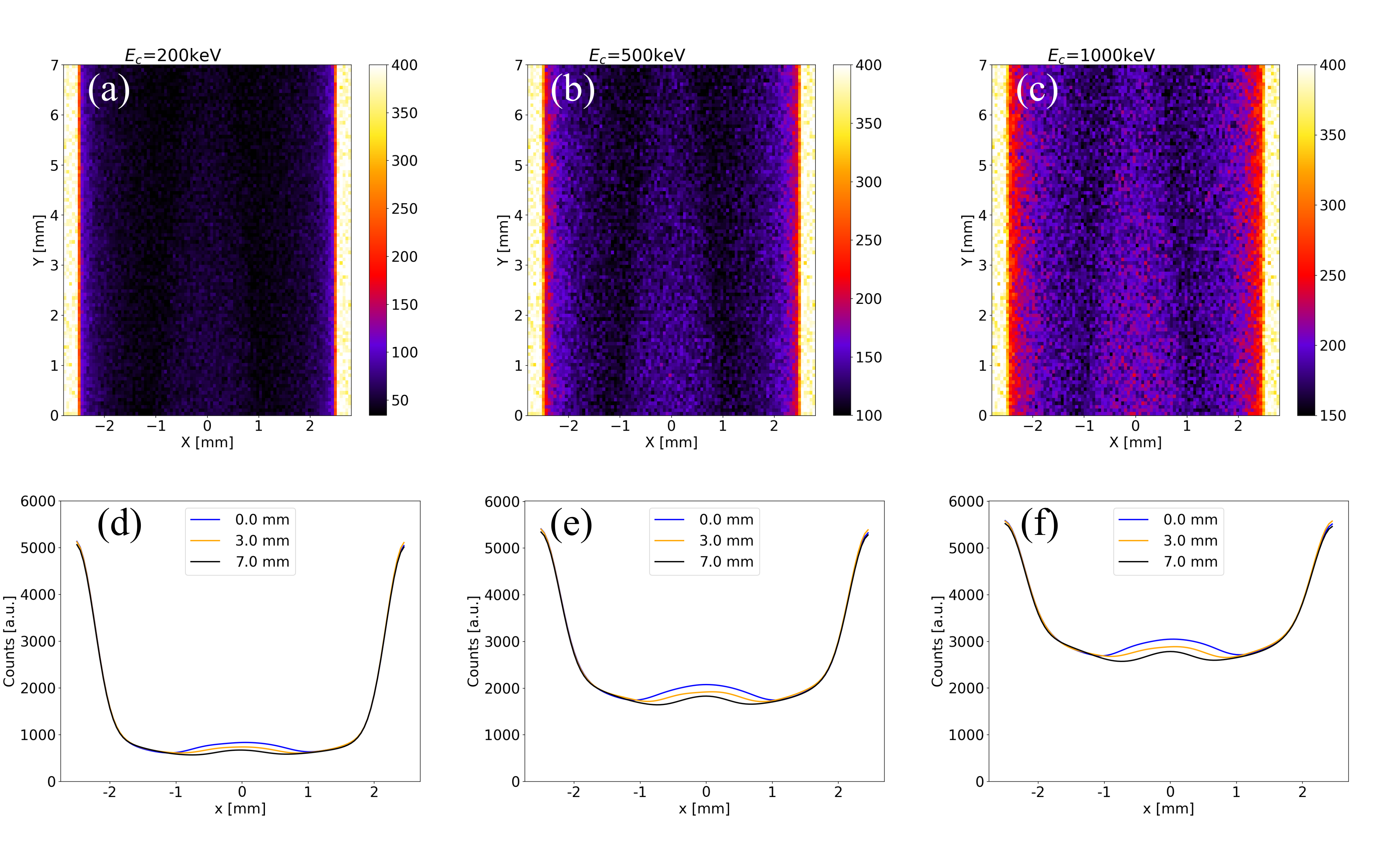}
\caption{
Simulated radiographs of the gas nozzle, performed with betatron radiation having $10^5 $ph/mm$^2$ and critical energies of (a) $E_c=200$ keV, (b) $E_c=500$ keV and (c) $E_c=1000$ keV. Their corresponding lineouts are shown in (d), (e), and (f).}\label{fig6}
\end{figure*}
	In Figs. 5(c) and 5(e), we compare lineouts taken at 3 different positions of the gas jet: $0$ mm (outlet), $3.5$ mm (middle cone region), and $7$ mm (inlet). The experimental lineout at the outlet (shown in Fig. 5(c)) has a signal intensity of $I_{sig} = 1930$ counts and a noise level $\sigma=51$ counts, giving a signal-to-noise ratio (SNR) of $I_{sig}/\sigma  \approx 38$, high enough to extract the inner dimensions. We note that the SNR ratios are comparable for both simulation and experiment. The small bumps around the center of the lineouts represent the inner structure through which gas flows. In Fig. 5(c), the bump (blue line, $0$ mm) extends from $-1$ mm to $+1$ mm, i.e., a diameter of $2$ mm, and is reproduced in the simulated lineout (Fig. 5(e), blue line). The same bump features are also in agreement for the lineout at $3.5$ mm (Fig. 5 (d) and Fig. 5(e), yellow line). The bump at the inlet location ($7$ mm) is visible in the simulation (Fig. 5(e), black line), while in the experiment, it can be extracted after accounting for the spatial variation of the signal (Fig.5 (c)). Note that the weak visibility is due to a low contrast-to-noise ratio (CNR) of the signal $(I_{sig} – I_{bg} )/\sigma = 2.6$ and the spatial variation of the betatron beam intensity. Such spatial effects are expected to occur, as the gas nozzle was located several mm away from the center of the beam, where angular effects reduce the photon number and energy \citep{bib39}. Overall, the investigation demonstrated that a single-shot radiograph can reveal the inlet size, outlet size, and opening angle of the gas jet, while the experimental data is in good agreement with simulations of the image and its lineouts.

	In Fig. 6, we examine further the influence of $\gamma$-ray energy on the properties of gas nozzle radiographs. GEANT4 simulations were performed using $10^5 $ph/mm$^2$, and the critical energy varied from $E_c=200 keV$ to $E_c=1000 keV$.  The radiographs (Figs. 6(a)-(c)) reveal that the increase in critical energy strongly improves the visibility of the inner structure of the nozzle. Figure 6(b) uses $500$ keV, and it is the image that shows the most similar contrast and magnitude with the experimental one (shown in Fig. 5), while for $200$ keV (Fig. 6(b)), the visibility is severely reduced due to a low transmissivity of the $\gamma$-rays. In all of the lineouts (Figs. 6(d)-(f)), we can observe the bump in transmissivity around the center of the lineout, corresponding to the inner structure of the nozzle. For the $200$ keV case, the magnitude of the bump is small, with $222$ counts at the outlet (Fig. 5(d) blue line). As we increase the energy from $E_c=200$ keV to $E_c=1000$ keV, the magnitude of the bump at the outlet lineout increases by $64\%$ while the magnitude of the bump at the inlet (black lines) increases more than $2\times$ (from $104$ to $215$ counts). This change significantly improves the signal and highlights that a high gamma energy extending in the MeV range is beneficial for such radiographs. On the other side, spectra with a low critical energy ($E_c=200$ keV ) would produce low-quality radiographs, making it impossible to identify the internal structure of the jet. Therefore, the dependence of the radiograph on the $\gamma$-ray energy serves as an additional evaluation for the betatron beam parameters, and we observe the highest similarity between simulated and experimental radiographs for the case of $E_c= 500 $ keV.

\section{Conclusion}\label{sec4}

The present work demonstrated the production and imaging application of high-energy betatron radiation ($E_c=515$ keV) with a 10-mrad divergence. The source was generated from LWFA-accelerated electron beams with energy up to 2 GeV (215 pC), produced by a 4-PW laser. For the present work, we developed an x-ray imaging setup operated at 0.1 Hz and which can work, in principle, at high-repetition rates (up to kHz), practically providing a spatial resolution of 38.5 $\mu$m/pixel. Our x-ray source and imaging setup have been successfully tested for non-destructive imaging of several cm-scale objects: a DRAM circuit, SMA and BNC connectors, a padlock, and a gas jet nozzle. The spatial resolution of this imaging can be as low as $150$ $\mu$m. The present work allowed us to identify the internal features of these objects, and as an example, we demonstrated the extraction of the inlet size, outlet size, and opening angle of a typical gas nozzle used in laser-plasma experiments. GEANT4 simulations of the radiography showed a good agreement with the experimental result, confirming the operating regime. The results demonstrate that betatron sources can reveal fine details of dense objects, ideal for non-destructive imaging in industry. Such a source can be useful for the detection of cracks and holes, soldering defects in 3D additive manufacturing, etc. Moreover, the high-energy range of the source spectrum, extending into the multi-MeV regime, and its ultrashort duration (fs-level) could prove useful for diagnosing high-density states of matter, such as those found in laser-driven fusion, without the need for additional targets for $\gamma$-ray production.

\bmhead{Acknowledgments}
The authors would like to acknowledge the operation of the PW laser and the target area management by the CoReLS technical staff. This work was supported by Institute for Basic Science under IBS-R012-D1.

\bibliography{sn-bibliography}


\begin{thebibliography}{46}
\ifx \bisbn   \undefined \def \bisbn  #1{ISBN #1}\fi
\ifx \binits  \undefined \def \binits#1{#1}\fi
\ifx \bauthor  \undefined \def \bauthor#1{#1}\fi
\ifx \batitle  \undefined \def \batitle#1{#1}\fi
\ifx \bjtitle  \undefined \def \bjtitle#1{#1}\fi
\ifx \bvolume  \undefined \def \bvolume#1{\textbf{#1}}\fi
\ifx \byear  \undefined \def \byear#1{#1}\fi
\ifx \bissue  \undefined \def \bissue#1{#1}\fi
\ifx \bfpage  \undefined \def \bfpage#1{#1}\fi
\ifx \blpage  \undefined \def \blpage #1{#1}\fi
\ifx \burl  \undefined \def \burl#1{\textsf{#1}}\fi
\ifx \doiurl  \undefined \def \doiurl#1{\url{https://doi.org/#1}}\fi
\ifx \betal  \undefined \def \betal{\textit{et al.}}\fi
\ifx \binstitute  \undefined \def \binstitute#1{#1}\fi
\ifx \binstitutionaled  \undefined \def \binstitutionaled#1{#1}\fi
\ifx \bctitle  \undefined \def \bctitle#1{#1}\fi
\ifx \beditor  \undefined \def \beditor#1{#1}\fi
\ifx \bpublisher  \undefined \def \bpublisher#1{#1}\fi
\ifx \bbtitle  \undefined \def \bbtitle#1{#1}\fi
\ifx \bedition  \undefined \def \bedition#1{#1}\fi
\ifx \bseriesno  \undefined \def \bseriesno#1{#1}\fi
\ifx \blocation  \undefined \def \blocation#1{#1}\fi
\ifx \bsertitle  \undefined \def \bsertitle#1{#1}\fi
\ifx \bsnm \undefined \def \bsnm#1{#1}\fi
\ifx \bsuffix \undefined \def \bsuffix#1{#1}\fi
\ifx \bparticle \undefined \def \bparticle#1{#1}\fi
\ifx \barticle \undefined \def \barticle#1{#1}\fi
\bibcommenthead
\ifx \bconfdate \undefined \def \bconfdate #1{#1}\fi
\ifx \botherref \undefined \def \botherref #1{#1}\fi
\ifx \url \undefined \def \url#1{\textsf{#1}}\fi
\ifx \bchapter \undefined \def \bchapter#1{#1}\fi
\ifx \bbook \undefined \def \bbook#1{#1}\fi
\ifx \bcomment \undefined \def \bcomment#1{#1}\fi
\ifx \oauthor \undefined \def \oauthor#1{#1}\fi
\ifx \citeauthoryear \undefined \def \citeauthoryear#1{#1}\fi
\ifx \endbibitem  \undefined \def \endbibitem {}\fi
\ifx \bconflocation  \undefined \def \bconflocation#1{#1}\fi
\ifx \arxivurl  \undefined \def \arxivurl#1{\textsf{#1}}\fi
\csname PreBibitemsHook\endcsname

\bibitem{bib1}
\begin{barticle}
\bauthor{\bsnm{Danson}, \binits{C.}},
\bauthor{\bsnm{Hillier}, \binits{D.}},
\bauthor{\bsnm{Hopps}, \binits{N.}},
\bauthor{\bsnm{Neely}, \binits{D.}}:
\batitle{Petawatt class lasers worldwide}.
\bjtitle{High Power Laser Sci. Eng.}
\bvolume{3},
\bfpage{3}
(\byear{2015}).
\doiurl{10.1017/hpl.2014.52}
\end{barticle}
\endbibitem

\bibitem{bib2}
\begin{barticle}
\bauthor{\bsnm{Esarey}, \binits{E.}},
\bauthor{\bsnm{Schroeder}, \binits{C.B.}},
\bauthor{\bsnm{Leemans}, \binits{W.P.}}:
\batitle{Physics of laser-driven plasma-based electron accelerators}.
\bjtitle{Rev. Mod. Phys.}
\bvolume{81},
\bfpage{1229}
(\byear{2009}).
\doiurl{10.1103/RevModPhys.81.1229}
\end{barticle}
\endbibitem

\bibitem{bib3}
\begin{barticle}
\bauthor{\bsnm{Downer}, \binits{M.C.}},
\bauthor{\bsnm{Zgadzaj}, \binits{R.}},
\bauthor{\bsnm{Debus}, \binits{A.}},
\bauthor{\bsnm{Schramm}, \binits{U.}},
\bauthor{\bsnm{Kaluza}, \binits{M.C.}}:
\batitle{Diagnostics for plasma-based electron accelerators}.
\bjtitle{Rev. Mod. Phys.}
\bvolume{90},
\bfpage{035002}
(\byear{2018}).
\doiurl{10.1103/RevModPhys.90.035002}
\end{barticle}
\endbibitem

\bibitem{bib4}
\begin{barticle}
\bauthor{\bsnm{Gonsalves}, \binits{A.J.}},
\bauthor{\bsnm{Nakamura}, \binits{K.}},
\bauthor{\bsnm{Daniels}, \binits{J.}},
\bauthor{\bsnm{Benedetti}, \binits{C.}},
\bauthor{\bsnm{Pieronek}, \binits{C.}},
\bauthor{\bparticle{de} \bsnm{Raadt}, \binits{T.C.H.}},
\bauthor{\bsnm{Steinke}, \binits{S.}},
\bauthor{\bsnm{Bin}, \binits{J.H.}},
\bauthor{\bsnm{Bulanov}, \binits{S.S.}},
\bauthor{\bparticle{van} \bsnm{Tilborg}, \binits{J.}},
\bauthor{\bsnm{Geddes}, \binits{C.G.R.}},
\bauthor{\bsnm{Schroeder}, \binits{C.B.}},
\bauthor{\bsnm{T\'oth}, \binits{C.}},
\bauthor{\bsnm{Esarey}, \binits{E.}},
\bauthor{\bsnm{Swanson}, \binits{K.}},
\bauthor{\bsnm{Fan-Chiang}, \binits{L.}},
\bauthor{\bsnm{Bagdasarov}, \binits{G.}},
\bauthor{\bsnm{Bobrova}, \binits{N.}},
\bauthor{\bsnm{Gasilov}, \binits{V.}},
\bauthor{\bsnm{Korn}, \binits{G.}},
\bauthor{\bsnm{Sasorov}, \binits{P.}},
\bauthor{\bsnm{Leemans}, \binits{W.P.}}:
\batitle{Petawatt Laser Guiding and Electron Beam Acceleration to 8 GeV in a
  Laser-Heated Capillary Discharge Waveguide}.
\bjtitle{Phys. Rev. Lett.}
\bvolume{122},
\bfpage{084801}
(\byear{2019}).
\doiurl{10.1103/PhysRevLett.122.084801}
\end{barticle}
\endbibitem

\bibitem{bib5}
\begin{botherref}
\oauthor{\bsnm{Aniculaesei}, \binits{C.}},
\oauthor{\bsnm{Ha}, \binits{T.}},
\oauthor{\bsnm{Yoffe}, \binits{S.}},
\oauthor{\bsnm{McCary}, \binits{E.}},
\oauthor{\bsnm{Spinks}, \binits{M.M.}},
\oauthor{\bsnm{Quevedo}, \binits{H.J.}},
\oauthor{\bsnm{Labun}, \binits{L.}},
\oauthor{\bsnm{Labun}, \binits{O.Z.}},
\oauthor{\bsnm{Sain}, \binits{R.}},
\oauthor{\bsnm{Hannasch}, \binits{A.}},
\oauthor{\bsnm{Zgadzaj}, \binits{R.}},
\oauthor{\bsnm{Pagano}, \binits{I.}},
\oauthor{\bsnm{Franco-Altamirano}, \binits{J.A.}},
\oauthor{\bsnm{Ringuette}, \binits{M.L.}},
\oauthor{\bsnm{Gaul}, \binits{E.}},
\oauthor{\bsnm{Luedtke}, \binits{S.V.}},
\oauthor{\bsnm{Tiwari}, \binits{G.}},
\oauthor{\bsnm{Ersfeld}, \binits{B.}},
\oauthor{\bsnm{Brunetti}, \binits{E.}},
\oauthor{\bsnm{Ruhl}, \binits{H.}},
\oauthor{\bsnm{Ditmire}, \binits{T.}},
\oauthor{\bsnm{Bruce}, \binits{S.}},
\oauthor{\bsnm{Donovan}, \binits{M.E.}},
\oauthor{\bsnm{Jaroszynski}, \binits{D.A.}},
\oauthor{\bsnm{Downer}, \binits{M.C.}},
\oauthor{\bsnm{Hegelich}, \binits{B.M.}}:
High-charge 10 GeV electron acceleration in a 10 cm nanoparticle-assisted
  hybrid wakefield accelerator
(2022).
\doiurl{10.48550/arXiv.2207.11492}
\end{botherref}
\endbibitem

\bibitem{bib6}
\begin{barticle}
\bauthor{\bsnm{Shaw}, \binits{J.L.}},
\bauthor{\bsnm{Romo-Gonzalez}, \binits{M.A.}},
\bauthor{\bsnm{Lemos}, \binits{N.}},
\bauthor{\bsnm{King}, \binits{P.M.}},
\bauthor{\bsnm{Bruhaug}, \binits{G.}},
\bauthor{\bsnm{Miller}, \binits{K.G.}},
\bauthor{\bsnm{Dorrer}, \binits{C.}},
\bauthor{\bsnm{Kruschwitz}, \binits{B.}},
\bauthor{\bsnm{Waxer}, \binits{L.}},
\bauthor{\bsnm{Williams}, \binits{G.J.}},
\bauthor{\bsnm{Ambat}, \binits{M.V.}},
\bauthor{\bsnm{McKie}, \binits{M.M.}},
\bauthor{\bsnm{Sinclair}, \binits{M.D.}},
\bauthor{\bsnm{Mori}, \binits{W.B.}},
\bauthor{\bsnm{Joshi}, \binits{C.}},
\bauthor{\bsnm{Chen}, \binits{H.}},
\bauthor{\bsnm{Palastro}, \binits{J.P.}},
\bauthor{\bsnm{Albert}, \binits{F.}},
\bauthor{\bsnm{Froula}, \binits{D.H.}}:
\batitle{Microcoulomb (0.7 $\pm \frac{0.4}{0.2} \mu$C) laser plasma accelerator
  on OMEGA EP}.
\bjtitle{Sci. Rep.}
\bvolume{11}(\bissue{1}),
\bfpage{7498}
(\byear{2021}).
\doiurl{10.1038/s41598-021-86523-5}
\end{barticle}
\endbibitem

\bibitem{bib7}
\begin{barticle}
\bauthor{\bsnm{Ke}, \binits{L.T.}},
\bauthor{\bsnm{Feng}, \binits{K.}},
\bauthor{\bsnm{Wang}, \binits{W.T.}},
\bauthor{\bsnm{Qin}, \binits{Z.Y.}},
\bauthor{\bsnm{Yu}, \binits{C.H.}},
\bauthor{\bsnm{Wu}, \binits{Y.}},
\bauthor{\bsnm{Chen}, \binits{Y.}},
\bauthor{\bsnm{Qi}, \binits{R.}},
\bauthor{\bsnm{Zhang}, \binits{Z.J.}},
\bauthor{\bsnm{Xu}, \binits{Y.}},
\bauthor{\bsnm{Yang}, \binits{X.J.}},
\bauthor{\bsnm{Leng}, \binits{Y.X.}},
\bauthor{\bsnm{Liu}, \binits{J.S.}},
\bauthor{\bsnm{Li}, \binits{R.X.}},
\bauthor{\bsnm{Xu}, \binits{Z.Z.}}:
\batitle{Near-GeV Electron Beams at a Few Per-Mille Level from a Laser
  Wakefield Accelerator via Density-Tailored Plasma}.
\bjtitle{Phys. Rev. Lett.}
\bvolume{126},
\bfpage{214801}
(\byear{2021}).
\doiurl{10.1103/PhysRevLett.126.214801}
\end{barticle}
\endbibitem

\bibitem{bib8}
\begin{barticle}
\bauthor{\bsnm{Gu{\'e}not}, \binits{D.}},
\bauthor{\bsnm{Gustas}, \binits{D.}},
\bauthor{\bsnm{Vernier}, \binits{A.}},
\bauthor{\bsnm{Beaurepaire}, \binits{B.}},
\bauthor{\bsnm{B{\"o}hle}, \binits{F.}},
\bauthor{\bsnm{Bocoum}, \binits{M.}},
\bauthor{\bsnm{Lozano}, \binits{M.}},
\bauthor{\bsnm{Jullien}, \binits{A.}},
\bauthor{\bsnm{Lopez-Martens}, \binits{R.}},
\bauthor{\bsnm{Lifschitz}, \binits{A.}},
\bauthor{\bsnm{Faure}, \binits{J.}}:
\batitle{Relativistic electron beams driven by kHz single-cycle light pulses}.
\bjtitle{Nat. Phot.}
\bvolume{11}(\bissue{5}),
\bfpage{293}
(\byear{2017}).
\doiurl{10.1038/nphoton.2017.46}
\end{barticle}
\endbibitem

\bibitem{bib9}
\begin{barticle}
\bauthor{\bsnm{Salehi}, \binits{F.}},
\bauthor{\bsnm{Goers}, \binits{A.J.}},
\bauthor{\bsnm{Hine}, \binits{G.A.}},
\bauthor{\bsnm{Feder}, \binits{L.}},
\bauthor{\bsnm{Kuk}, \binits{D.}},
\bauthor{\bsnm{Miao}, \binits{B.}},
\bauthor{\bsnm{Woodbury}, \binits{D.}},
\bauthor{\bsnm{Kim}, \binits{K.Y.}},
\bauthor{\bsnm{Milchberg}, \binits{H.M.}}:
\batitle{MeV electron acceleration at 1 kHz with $\ll 10$ mJ laser pulses}.
\bjtitle{Opt. Lett.}
\bvolume{42}(\bissue{2}),
\bfpage{215}
(\byear{2017}).
\doiurl{10.1364/OL.42.000215}
\end{barticle}
\endbibitem

\bibitem{bib10}
\begin{barticle}
\bauthor{\bsnm{Albert}, \binits{F.}},
\bauthor{\bsnm{Thomas}, \binits{A.G.R.}}:
\batitle{Applications of laser wakefield accelerator-based light sources}.
\bjtitle{Plasma Phys. Controlled Fusion}
\bvolume{58}(\bissue{10}),
\bfpage{103001}
(\byear{2016}).
\doiurl{10.1088/0741-3335/58/10/103001}
\end{barticle}
\endbibitem

\bibitem{bib11}
\begin{barticle}
\bauthor{\bsnm{Wang}, \binits{W.}},
\bauthor{\bsnm{Feng}, \binits{K.}},
\bauthor{\bsnm{Ke}, \binits{L.}},
\bauthor{\bsnm{Yu}, \binits{C.}},
\bauthor{\bsnm{Xu}, \binits{Y.}},
\bauthor{\bsnm{Qi}, \binits{R.}},
\bauthor{\bsnm{Chen}, \binits{Y.}},
\bauthor{\bsnm{Qin}, \binits{Z.}},
\bauthor{\bsnm{Zhang}, \binits{Z.}},
\bauthor{\bsnm{Fang}, \binits{M.}},
\bauthor{\bsnm{Liu}, \binits{J.}},
\bauthor{\bsnm{Jiang}, \binits{K.}},
\bauthor{\bsnm{Wang}, \binits{H.}},
\bauthor{\bsnm{Wang}, \binits{C.}},
\bauthor{\bsnm{Yang}, \binits{X.}},
\bauthor{\bsnm{Wu}, \binits{F.}},
\bauthor{\bsnm{Leng}, \binits{Y.}},
\bauthor{\bsnm{Liu}, \binits{J.}},
\bauthor{\bsnm{Li}, \binits{R.}},
\bauthor{\bsnm{Xu}, \binits{Z.}}:
\batitle{Free-electron lasing at 27 nanometres based on a laser wakefield
  accelerator}.
\bjtitle{Nature}
\bvolume{595}(\bissue{7868}),
\bfpage{516}
(\byear{2021}).
\doiurl{10.1038/s41586-021-03678-x}
\end{barticle}
\endbibitem

\bibitem{bib12}
\begin{barticle}
\bauthor{\bsnm{Chen}, \binits{S.}},
\bauthor{\bsnm{Powers}, \binits{N.D.}},
\bauthor{\bsnm{Ghebregziabher}, \binits{I.}},
\bauthor{\bsnm{Maharjan}, \binits{C.M.}},
\bauthor{\bsnm{Liu}, \binits{C.}},
\bauthor{\bsnm{Golovin}, \binits{G.}},
\bauthor{\bsnm{Banerjee}, \binits{S.}},
\bauthor{\bsnm{Zhang}, \binits{J.}},
\bauthor{\bsnm{Cunningham}, \binits{N.}},
\bauthor{\bsnm{Moorti}, \binits{A.}},
\bauthor{\bsnm{Clarke}, \binits{S.}},
\bauthor{\bsnm{Pozzi}, \binits{S.}},
\bauthor{\bsnm{Umstadter}, \binits{D.P.}}:
\batitle{MeV-Energy X-Rays from Inverse Compton Scattering with Laser-Wakefield
  Accelerated Electrons}.
\bjtitle{Phys. Rev. Lett.}
\bvolume{110},
\bfpage{155003}
(\byear{2013}).
\doiurl{10.1103/PhysRevLett.110.155003}
\end{barticle}
\endbibitem

\bibitem{bib13}
\begin{barticle}
\bauthor{\bsnm{Sarri}, \binits{G.}},
\bauthor{\bsnm{Corvan}, \binits{D.J.}},
\bauthor{\bsnm{Schumaker}, \binits{W.}},
\bauthor{\bsnm{Cole}, \binits{J.M.}},
\bauthor{\bsnm{Di~Piazza}, \binits{A.}},
\bauthor{\bsnm{Ahmed}, \binits{H.}},
\bauthor{\bsnm{Harvey}, \binits{C.}},
\bauthor{\bsnm{Keitel}, \binits{C.H.}},
\bauthor{\bsnm{Krushelnick}, \binits{K.}},
\bauthor{\bsnm{Mangles}, \binits{S.P.D.}},
\bauthor{\bsnm{Najmudin}, \binits{Z.}},
\bauthor{\bsnm{Symes}, \binits{D.}},
\bauthor{\bsnm{Thomas}, \binits{A.G.R.}},
\bauthor{\bsnm{Yeung}, \binits{M.}},
\bauthor{\bsnm{Zhao}, \binits{Z.}},
\bauthor{\bsnm{Zepf}, \binits{M.}}:
\batitle{Ultrahigh Brilliance Multi-MeV $\ensuremath{\gamma}$-Ray Beams from
  Nonlinear Relativistic Thomson Scattering}.
\bjtitle{Phys. Rev. Lett.}
\bvolume{113},
\bfpage{224801}
(\byear{2014}).
\doiurl{10.1103/PhysRevLett.113.224801}
\end{barticle}
\endbibitem

\bibitem{bib14}
\begin{barticle}
\bauthor{\bsnm{Rousse}, \binits{A.}},
\bauthor{\bsnm{Phuoc}, \binits{K.T.}},
\bauthor{\bsnm{Shah}, \binits{R.}},
\bauthor{\bsnm{Pukhov}, \binits{A.}},
\bauthor{\bsnm{Lefebvre}, \binits{E.}},
\bauthor{\bsnm{Malka}, \binits{V.}},
\bauthor{\bsnm{Kiselev}, \binits{S.}},
\bauthor{\bsnm{Burgy}, \binits{F.}},
\bauthor{\bsnm{Rousseau}, \binits{J.-P.}},
\bauthor{\bsnm{Umstadter}, \binits{D.}},
\bauthor{\bsnm{Hulin}, \binits{D.}}:
\batitle{Production of a keV X-Ray Beam from Synchrotron Radiation in
  Relativistic Laser-Plasma Interaction}.
\bjtitle{Phys. Rev. Lett.}
\bvolume{93},
\bfpage{135005}
(\byear{2004}).
\doiurl{10.1103/PhysRevLett.93.135005}
\end{barticle}
\endbibitem

\bibitem{bib15}
\begin{barticle}
\bauthor{\bsnm{Esarey}, \binits{E.}},
\bauthor{\bsnm{Shadwick}, \binits{B.A.}},
\bauthor{\bsnm{Catravas}, \binits{P.}},
\bauthor{\bsnm{Leemans}, \binits{W.P.}}:
\batitle{Synchrotron radiation from electron beams in plasma-focusing
  channels}.
\bjtitle{Phys. Rev. E}
\bvolume{65},
\bfpage{056505}
(\byear{2002}).
\doiurl{10.1103/PhysRevE.65.056505}
\end{barticle}
\endbibitem

\bibitem{bib16}
\begin{barticle}
\bauthor{\bsnm{Corde}, \binits{S.}},
\bauthor{\bsnm{Ta~Phuoc}, \binits{K.}},
\bauthor{\bsnm{Lambert}, \binits{G.}},
\bauthor{\bsnm{Fitour}, \binits{R.}},
\bauthor{\bsnm{Malka}, \binits{V.}},
\bauthor{\bsnm{Rousse}, \binits{A.}},
\bauthor{\bsnm{Beck}, \binits{A.}},
\bauthor{\bsnm{Lefebvre}, \binits{E.}}:
\batitle{Femtosecond x rays from laser-plasma accelerators}.
\bjtitle{Rev. Mod. Phys.}
\bvolume{85},
\bfpage{1}
(\byear{2013}).
\doiurl{10.1103/RevModPhys.85.1}
\end{barticle}
\endbibitem

\bibitem{bib17}
\begin{barticle}
\bauthor{\bsnm{Hussein}, \binits{A.E.}},
\bauthor{\bsnm{Senabulya}, \binits{N.}},
\bauthor{\bsnm{Ma}, \binits{Y.}},
\bauthor{\bsnm{Streeter}, \binits{M.J.V.}},
\bauthor{\bsnm{Kettle}, \binits{B.}},
\bauthor{\bsnm{Dann}, \binits{S.J.D.}},
\bauthor{\bsnm{Albert}, \binits{F.}},
\bauthor{\bsnm{Bourgeois}, \binits{N.}},
\bauthor{\bsnm{Cipiccia}, \binits{S.}},
\bauthor{\bsnm{Cole}, \binits{J.M.}},
\bauthor{\bsnm{Finlay}, \binits{O.}},
\bauthor{\bsnm{Gerstmayr}, \binits{E.}},
\bauthor{\bsnm{Gonz{\'a}lez}, \binits{I.G.}},
\bauthor{\bsnm{Higginbotham}, \binits{A.}},
\bauthor{\bsnm{Jaroszynski}, \binits{D.A.}},
\bauthor{\bsnm{Falk}, \binits{K.}},
\bauthor{\bsnm{Krushelnick}, \binits{K.}},
\bauthor{\bsnm{Lemos}, \binits{N.}},
\bauthor{\bsnm{Lopes}, \binits{N.C.}},
\bauthor{\bsnm{Lumsdon}, \binits{C.}},
\bauthor{\bsnm{Lundh}, \binits{O.}},
\bauthor{\bsnm{Mangles}, \binits{S.P.D.}},
\bauthor{\bsnm{Najmudin}, \binits{Z.}},
\bauthor{\bsnm{Rajeev}, \binits{P.P.}},
\bauthor{\bsnm{Schlep{\"u}tz}, \binits{C.M.}},
\bauthor{\bsnm{Shahzad}, \binits{M.}},
\bauthor{\bsnm{Smid}, \binits{M.}},
\bauthor{\bsnm{Spesyvtsev}, \binits{R.}},
\bauthor{\bsnm{Symes}, \binits{D.R.}},
\bauthor{\bsnm{Vieux}, \binits{G.}},
\bauthor{\bsnm{Willingale}, \binits{L.}},
\bauthor{\bsnm{Wood}, \binits{J.C.}},
\bauthor{\bsnm{Shahani}, \binits{A.J.}},
\bauthor{\bsnm{Thomas}, \binits{A.G.R.}}:
\batitle{Laser-wakefield accelerators for high-resolution X-ray imaging of
  complex microstructures}.
\bjtitle{Sci. Rep.}
\bvolume{9}(\bissue{1}),
\bfpage{3249}
(\byear{2019}).
\doiurl{10.1038/s41598-019-39845-4}
\end{barticle}
\endbibitem

\bibitem{bib18}
\begin{barticle}
\bauthor{\bsnm{Fourmaux}, \binits{S.}},
\bauthor{\bsnm{Corde}, \binits{S.}},
\bauthor{\bsnm{Phuoc}, \binits{K.T.}},
\bauthor{\bsnm{Lassonde}, \binits{P.}},
\bauthor{\bsnm{Lebrun}, \binits{G.}},
\bauthor{\bsnm{Payeur}, \binits{S.}},
\bauthor{\bsnm{Martin}, \binits{F.}},
\bauthor{\bsnm{Sebban}, \binits{S.}},
\bauthor{\bsnm{Malka}, \binits{V.}},
\bauthor{\bsnm{Rousse}, \binits{A.}},
\bauthor{\bsnm{Kieffer}, \binits{J.C.}}:
\batitle{Single shot phase contrast imaging using laser-produced Betatron x-ray
  beams}.
\bjtitle{Opt. Lett.}
\bvolume{36}(\bissue{13}),
\bfpage{2426}
(\byear{2011}).
\doiurl{10.1364/OL.36.002426}
\end{barticle}
\endbibitem

\bibitem{bib19}
\begin{barticle}
\bauthor{\bsnm{Kneip}, \binits{S.}},
\bauthor{\bsnm{McGuffey}, \binits{C.}},
\bauthor{\bsnm{Dollar}, \binits{F.}},
\bauthor{\bsnm{Bloom}, \binits{M.S.}},
\bauthor{\bsnm{Chvykov}, \binits{V.}},
\bauthor{\bsnm{Kalintchenko}, \binits{G.}},
\bauthor{\bsnm{Krushelnick}, \binits{K.}},
\bauthor{\bsnm{Maksimchuk}, \binits{A.}},
\bauthor{\bsnm{Mangles}, \binits{S.P.D.}},
\bauthor{\bsnm{Matsuoka}, \binits{T.}},
\bauthor{\bsnm{Najmudin}, \binits{Z.}},
\bauthor{\bsnm{Palmer}, \binits{C.A.J.}},
\bauthor{\bsnm{Schreiber}, \binits{J.}},
\bauthor{\bsnm{Schumaker}, \binits{W.}},
\bauthor{\bsnm{Thomas}, \binits{A.G.R.}},
\bauthor{\bsnm{Yanovsky}, \binits{V.}}:
\batitle{X-ray phase contrast imaging of biological specimens with femtosecond
  pulses of betatron radiation from a compact laser plasma wakefield
  accelerator}.
\bjtitle{Applied Phys. Lett.}
\bvolume{99}(\bissue{9}),
\bfpage{093701}
(\byear{2011})
{\href{https://arxiv.org/abs/https://doi.org/10.1063/1.3627216}{{https://doi.org/10.1063/1.3627216}}}.
\doiurl{10.1063/1.3627216}
\end{barticle}
\endbibitem

\bibitem{bib20}
\begin{barticle}
\bauthor{\bsnm{Wenz}, \binits{J.}},
\bauthor{\bsnm{Schleede}, \binits{S.}},
\bauthor{\bsnm{Khrennikov}, \binits{K.}},
\bauthor{\bsnm{Bech}, \binits{M.}},
\bauthor{\bsnm{Thibault}, \binits{P.}},
\bauthor{\bsnm{Heigoldt}, \binits{M.}},
\bauthor{\bsnm{Pfeiffer}, \binits{F.}},
\bauthor{\bsnm{Karsch}, \binits{S.}}:
\batitle{Quantitative X-ray phase-contrast microtomography from a compact
  laser-driven betatron source}.
\bjtitle{Nat. Comm.}
\bvolume{6}(\bissue{1}),
\bfpage{7568}
(\byear{2015}).
\doiurl{10.1038/ncomms8568}
\end{barticle}
\endbibitem

\bibitem{bib21}
\begin{barticle}
\bauthor{\bsnm{Cole}, \binits{J.M.}},
\bauthor{\bsnm{Wood}, \binits{J.C.}},
\bauthor{\bsnm{Lopes}, \binits{N.C.}},
\bauthor{\bsnm{Poder}, \binits{K.}},
\bauthor{\bsnm{Abel}, \binits{R.L.}},
\bauthor{\bsnm{Alatabi}, \binits{S.}},
\bauthor{\bsnm{Bryant}, \binits{J.S.J.}},
\bauthor{\bsnm{Jin}, \binits{A.}},
\bauthor{\bsnm{Kneip}, \binits{S.}},
\bauthor{\bsnm{Mecseki}, \binits{K.}},
\bauthor{\bsnm{Symes}, \binits{D.R.}},
\bauthor{\bsnm{Mangles}, \binits{S.P.D.}},
\bauthor{\bsnm{Najmudin}, \binits{Z.}}:
\batitle{Laser-wakefield accelerators as hard x-ray sources for 3D medical
  imaging of human bone}.
\bjtitle{Sci. Rep.}
\bvolume{5}(\bissue{1}),
\bfpage{13244}
(\byear{2015}).
\doiurl{10.1038/srep13244}
\end{barticle}
\endbibitem

\bibitem{bib22}
\begin{barticle}
\bauthor{\bsnm{Wood}, \binits{J.C.}},
\bauthor{\bsnm{Chapman}, \binits{D.J.}},
\bauthor{\bsnm{Poder}, \binits{K.}},
\bauthor{\bsnm{Lopes}, \binits{N.C.}},
\bauthor{\bsnm{Rutherford}, \binits{M.E.}},
\bauthor{\bsnm{White}, \binits{T.G.}},
\bauthor{\bsnm{Albert}, \binits{F.}},
\bauthor{\bsnm{Behm}, \binits{K.T.}},
\bauthor{\bsnm{Booth}, \binits{N.}},
\bauthor{\bsnm{Bryant}, \binits{J.S.J.}},
\bauthor{\bsnm{Foster}, \binits{P.S.}},
\bauthor{\bsnm{Glenzer}, \binits{S.}},
\bauthor{\bsnm{Hill}, \binits{E.}},
\bauthor{\bsnm{Krushelnick}, \binits{K.}},
\bauthor{\bsnm{Najmudin}, \binits{Z.}},
\bauthor{\bsnm{Pollock}, \binits{B.B.}},
\bauthor{\bsnm{Rose}, \binits{S.}},
\bauthor{\bsnm{Schumaker}, \binits{W.}},
\bauthor{\bsnm{Scott}, \binits{R.H.H.}},
\bauthor{\bsnm{Sherlock}, \binits{M.}},
\bauthor{\bsnm{Thomas}, \binits{A.G.R.}},
\bauthor{\bsnm{Zhao}, \binits{Z.}},
\bauthor{\bsnm{Eakins}, \binits{D.E.}},
\bauthor{\bsnm{Mangles}, \binits{S.P.D.}}:
\batitle{Ultrafast Imaging of Laser Driven Shock Waves using Betatron X-rays
  from a Laser Wakefield Accelerator}.
\bjtitle{Scie. Rep.}
\bvolume{8}(\bissue{1}),
\bfpage{11010}
(\byear{2018}).
\doiurl{10.1038/s41598-018-29347-0}
\end{barticle}
\endbibitem

\bibitem{bib23}
\begin{barticle}
\bauthor{\bsnm{Albert}, \binits{F.}},
\bauthor{\bsnm{Lemos}, \binits{N.}},
\bauthor{\bsnm{Shaw}, \binits{J.L.}},
\bauthor{\bsnm{King}, \binits{P.M.}},
\bauthor{\bsnm{Pollock}, \binits{B.B.}},
\bauthor{\bsnm{Goyon}, \binits{C.}},
\bauthor{\bsnm{Schumaker}, \binits{W.}},
\bauthor{\bsnm{Saunders}, \binits{A.M.}},
\bauthor{\bsnm{Marsh}, \binits{K.A.}},
\bauthor{\bsnm{Pak}, \binits{A.}},
\bauthor{\bsnm{Ralph}, \binits{J.E.}},
\bauthor{\bsnm{Martins}, \binits{J.L.}},
\bauthor{\bsnm{Amorim}, \binits{L.D.}},
\bauthor{\bsnm{Falcone}, \binits{R.W.}},
\bauthor{\bsnm{Glenzer}, \binits{S.H.}},
\bauthor{\bsnm{Moody}, \binits{J.D.}},
\bauthor{\bsnm{Joshi}, \binits{C.}}:
\batitle{Betatron x-ray radiation in the self-modulated laser wakefield
  acceleration regime: prospects for a novel probe at large scale laser
  facilities}.
\bjtitle{Nuclear Fusion}
\bvolume{59}(\bissue{3}),
\bfpage{032003}
(\byear{2018}).
\doiurl{10.1088/1741-4326/aad058}
\end{barticle}
\endbibitem

\bibitem{bib24}
\begin{barticle}
\bauthor{\bsnm{Gruse}, \binits{J.-N.}},
\bauthor{\bsnm{Streeter}, \binits{M.J.V.}},
\bauthor{\bsnm{Thornton}, \binits{C.}},
\bauthor{\bsnm{Armstrong}, \binits{C.D.}},
\bauthor{\bsnm{Baird}, \binits{C.D.}},
\bauthor{\bsnm{Bourgeois}, \binits{N.}},
\bauthor{\bsnm{Cipiccia}, \binits{S.}},
\bauthor{\bsnm{Finlay}, \binits{O.J.}},
\bauthor{\bsnm{Gregory}, \binits{C.D.}},
\bauthor{\bsnm{Katzir}, \binits{Y.}},
\bauthor{\bsnm{Lopes}, \binits{N.C.}},
\bauthor{\bsnm{Mangles}, \binits{S.P.D.}},
\bauthor{\bsnm{Najmudin}, \binits{Z.}},
\bauthor{\bsnm{Neely}, \binits{D.}},
\bauthor{\bsnm{Pickard}, \binits{L.R.}},
\bauthor{\bsnm{Potter}, \binits{K.D.}},
\bauthor{\bsnm{Rajeev}, \binits{P.P.}},
\bauthor{\bsnm{Rusby}, \binits{D.R.}},
\bauthor{\bsnm{Underwood}, \binits{C.I.D.}},
\bauthor{\bsnm{Warnett}, \binits{J.M.}},
\bauthor{\bsnm{Williams}, \binits{M.A.}},
\bauthor{\bsnm{Wood}, \binits{J.C.}},
\bauthor{\bsnm{Murphy}, \binits{C.D.}},
\bauthor{\bsnm{Brenner}, \binits{C.M.}},
\bauthor{\bsnm{Symes}, \binits{D.R.}}:
\batitle{Application of compact laser-driven accelerator X-ray sources for
  industrial imaging}.
\bjtitle{Nuclear Instruments and Methods in Physics Research Section A:
  Accelerators, Spectrometers, Detectors and Associated Equipment}
\bvolume{983},
\bfpage{164369}
(\byear{2020}).
\doiurl{10.1016/j.nima.2020.164369}
\end{barticle}
\endbibitem

\bibitem{bib25}
\begin{barticle}
\bauthor{\bsnm{Glinec}, \binits{Y.}},
\bauthor{\bsnm{Faure}, \binits{J.}},
\bauthor{\bsnm{Dain}, \binits{L.L.}},
\bauthor{\bsnm{Darbon}, \binits{S.}},
\bauthor{\bsnm{Hosokai}, \binits{T.}},
\bauthor{\bsnm{Santos}, \binits{J.J.}},
\bauthor{\bsnm{Lefebvre}, \binits{E.}},
\bauthor{\bsnm{Rousseau}, \binits{J.P.}},
\bauthor{\bsnm{Burgy}, \binits{F.}},
\bauthor{\bsnm{Mercier}, \binits{B.}},
\bauthor{\bsnm{Malka}, \binits{V.}}:
\batitle{High-Resolution $\ensuremath{\gamma}$-Ray Radiography Produced by a
  Laser-Plasma Driven Electron Source}.
\bjtitle{Phys. Rev. Lett.}
\bvolume{94},
\bfpage{025003}
(\byear{2005}).
\doiurl{10.1103/PhysRevLett.94.025003}
\end{barticle}
\endbibitem

\bibitem{bib26}
\begin{barticle}
\bauthor{\bsnm{Albert}, \binits{F.}},
\bauthor{\bsnm{Couprie}, \binits{M.E.}},
\bauthor{\bsnm{Debus}, \binits{A.}},
\bauthor{\bsnm{Downer}, \binits{M.C.}},
\bauthor{\bsnm{Faure}, \binits{J.}},
\bauthor{\bsnm{Flacco}, \binits{A.}},
\bauthor{\bsnm{Gizzi}, \binits{L.A.}},
\bauthor{\bsnm{Grismayer}, \binits{T.}},
\bauthor{\bsnm{Huebl}, \binits{A.}},
\bauthor{\bsnm{Joshi}, \binits{C.}},
\bauthor{\bsnm{Labat}, \binits{M.}},
\bauthor{\bsnm{Leemans}, \binits{W.P.}},
\bauthor{\bsnm{Maier}, \binits{A.R.}},
\bauthor{\bsnm{Mangles}, \binits{S.P.D.}},
\bauthor{\bsnm{Mason}, \binits{P.}},
\bauthor{\bsnm{Mathieu}, \binits{F.}},
\bauthor{\bsnm{Muggli}, \binits{P.}},
\bauthor{\bsnm{Nishiuchi}, \binits{M.}},
\bauthor{\bsnm{Osterhoff}, \binits{J.}},
\bauthor{\bsnm{Rajeev}, \binits{P.P.}},
\bauthor{\bsnm{Schramm}, \binits{U.}},
\bauthor{\bsnm{Schreiber}, \binits{J.}},
\bauthor{\bsnm{Thomas}, \binits{A.G.R.}},
\bauthor{\bsnm{Vay}, \binits{J.-L.}},
\bauthor{\bsnm{Vranic}, \binits{M.}},
\bauthor{\bsnm{Zeil}, \binits{K.}}:
\batitle{2020 roadmap on plasma accelerators}.
\bjtitle{New J. of Phys.}
\bvolume{23}(\bissue{3}),
\bfpage{031101}
(\byear{2021}).
\doiurl{10.1088/1367-2630/abcc62}
\end{barticle}
\endbibitem

\bibitem{bib27}
\begin{barticle}
\bauthor{\bsnm{Sung}, \binits{J.H.}},
\bauthor{\bsnm{Lee}, \binits{H.W.}},
\bauthor{\bsnm{Yoo}, \binits{J.Y.}},
\bauthor{\bsnm{Yoon}, \binits{J.W.}},
\bauthor{\bsnm{Lee}, \binits{C.W.}},
\bauthor{\bsnm{Yang}, \binits{J.M.}},
\bauthor{\bsnm{Son}, \binits{Y.J.}},
\bauthor{\bsnm{Jang}, \binits{Y.H.}},
\bauthor{\bsnm{Lee}, \binits{S.K.}},
\bauthor{\bsnm{Nam}, \binits{C.H.}}:
\batitle{4.2 PW 20 fs Ti:sapphire laser at 0.1 Hz}.
\bjtitle{Opt. Lett.}
\bvolume{42}(\bissue{11}),
\bfpage{2058}
(\byear{2017}).
\doiurl{10.1364/OL.42.002058}
\end{barticle}
\endbibitem

\bibitem{bib28}
\begin{barticle}
\bauthor{\bsnm{Clayton}, \binits{C.E.}},
\bauthor{\bsnm{Ralph}, \binits{J.E.}},
\bauthor{\bsnm{Albert}, \binits{F.}},
\bauthor{\bsnm{Fonseca}, \binits{R.A.}},
\bauthor{\bsnm{Glenzer}, \binits{S.H.}},
\bauthor{\bsnm{Joshi}, \binits{C.}},
\bauthor{\bsnm{Lu}, \binits{W.}},
\bauthor{\bsnm{Marsh}, \binits{K.A.}},
\bauthor{\bsnm{Martins}, \binits{S.F.}},
\bauthor{\bsnm{Mori}, \binits{W.B.}},
\bauthor{\bsnm{Pak}, \binits{A.}},
\bauthor{\bsnm{Tsung}, \binits{F.S.}},
\bauthor{\bsnm{Pollock}, \binits{B.B.}},
\bauthor{\bsnm{Ross}, \binits{J.S.}},
\bauthor{\bsnm{Silva}, \binits{L.O.}},
\bauthor{\bsnm{Froula}, \binits{D.H.}}:
\batitle{Self-Guided Laser Wakefield Acceleration beyond 1 GeV Using
  Ionization-Induced Injection}.
\bjtitle{Phys. Rev. Lett.}
\bvolume{105},
\bfpage{105003}
(\byear{2010}).
\doiurl{10.1103/PhysRevLett.105.105003}
\end{barticle}
\endbibitem

\bibitem{bib29}
\begin{barticle}
\bauthor{\bsnm{Mirzaie}, \binits{M.}},
\bauthor{\bsnm{Li}, \binits{S.}},
\bauthor{\bsnm{Zeng}, \binits{M.}},
\bauthor{\bsnm{Hafz}, \binits{N.A.M.}},
\bauthor{\bsnm{Chen}, \binits{M.}},
\bauthor{\bsnm{Li}, \binits{G.Y.}},
\bauthor{\bsnm{Zhu}, \binits{Q.J.}},
\bauthor{\bsnm{Liao}, \binits{H.}},
\bauthor{\bsnm{Sokollik}, \binits{T.}},
\bauthor{\bsnm{Liu}, \binits{F.}},
\bauthor{\bsnm{Ma}, \binits{Y.Y.}},
\bauthor{\bsnm{Chen}, \binits{L.M.}},
\bauthor{\bsnm{Sheng}, \binits{Z.M.}},
\bauthor{\bsnm{Zhang}, \binits{J.}}:
\batitle{Demonstration of self-truncated ionization injection for GeV electron
  beams}.
\bjtitle{Sci. Rep.}
\bvolume{5}(\bissue{1}),
\bfpage{14659}
(\byear{2015}).
\doiurl{10.1038/srep14659}
\end{barticle}
\endbibitem

\bibitem{bib30}
\begin{barticle}
\bauthor{\bsnm{Kim}, \binits{H.T.}},
\bauthor{\bsnm{Pathak}, \binits{V.B.}},
\bauthor{\bsnm{Hong~Pae}, \binits{K.}},
\bauthor{\bsnm{Lifschitz}, \binits{A.}},
\bauthor{\bsnm{Sylla}, \binits{F.}},
\bauthor{\bsnm{Shin}, \binits{J.H.}},
\bauthor{\bsnm{Hojbota}, \binits{C.}},
\bauthor{\bsnm{Lee}, \binits{S.K.}},
\bauthor{\bsnm{Sung}, \binits{J.H.}},
\bauthor{\bsnm{Lee}, \binits{H.W.}},
\bauthor{\bsnm{Guillaume}, \binits{E.}},
\bauthor{\bsnm{Thaury}, \binits{C.}},
\bauthor{\bsnm{Nakajima}, \binits{K.}},
\bauthor{\bsnm{Vieira}, \binits{J.}},
\bauthor{\bsnm{Silva}, \binits{L.O.}},
\bauthor{\bsnm{Malka}, \binits{V.}},
\bauthor{\bsnm{Nam}, \binits{C.H.}}:
\batitle{Stable multi-GeV electron accelerator driven by waveform-controlled PW
  laser pulses}.
\bjtitle{Sci. Rep.}
\bvolume{7}(\bissue{1}),
\bfpage{10203}
(\byear{2017}).
\doiurl{10.1038/s41598-017-09267-1}
\end{barticle}
\endbibitem

\bibitem{bib31}
\begin{barticle}
\bauthor{\bsnm{Hojbota}, \binits{C.I.}},
\bauthor{\bsnm{Kim}, \binits{H.T.}},
\bauthor{\bsnm{Shin}, \binits{J.H.}},
\bauthor{\bsnm{Aniculaesei}, \binits{C.}},
\bauthor{\bsnm{Rao}, \binits{B.S.}},
\bauthor{\bsnm{Nam}, \binits{C.H.}}:
\batitle{Accurate single-shot measurement technique for the spectral
  distribution of GeV electron beams from a laser wakefield accelerator}.
\bjtitle{AIP Advances}
\bvolume{9}(\bissue{8}),
\bfpage{085229}
(\byear{2019}).
\doiurl{10.1063/1.5117311}
\end{barticle}
\endbibitem

\bibitem{bib32}
\begin{botherref}
\oauthor{\bsnm{Pollock}, \binits{B.B.}},
\oauthor{\bsnm{Ross}, \binits{J.S.}},
\oauthor{\bsnm{Tynan}, \binits{G.R.}},
\oauthor{\bsnm{Divol}, \binits{L.}},
\oauthor{\bsnm{Glenzer}, \binits{S.H.}},
\oauthor{\bsnm{Leurent}, \binits{V.}},
\oauthor{\bsnm{Palastro}, \binits{J.P.}},
\oauthor{\bsnm{Ralph}, \binits{J.E.}},
\oauthor{\bsnm{Froula}, \binits{D.H.}},
\oauthor{\bsnm{Clayton}, \binits{C.E.}},
\oauthor{\bsnm{Marsh}, \binits{K.A.}},
\oauthor{\bsnm{Pak}, \binits{A.E.}},
\oauthor{\bsnm{Wang}, \binits{T.L.}},
\oauthor{\bsnm{Joshi}, \binits{C.}}:
Two-Screen Method for Determining Electron Beam Energy and Deflection from
  Laser Wakefield Acceleration.
Technical report,
United States
(Apr 2009).
LLNL-PROC--412609.
\url{https://e-reports-ext.llnl.gov/pdf/372646.pdf}
\end{botherref}
\endbibitem

\bibitem{bib33}
\begin{barticle}
\bauthor{\bsnm{Tanaka}, \binits{K.A.}},
\bauthor{\bsnm{Yabuuchi}, \binits{T.}},
\bauthor{\bsnm{Sato}, \binits{T.}},
\bauthor{\bsnm{Kodama}, \binits{R.}},
\bauthor{\bsnm{Kitagawa}, \binits{Y.}},
\bauthor{\bsnm{Takahashi}, \binits{T.}},
\bauthor{\bsnm{Ikeda}, \binits{T.}},
\bauthor{\bsnm{Honda}, \binits{Y.}},
\bauthor{\bsnm{Okuda}, \binits{S.}}:
\batitle{Calibration of imaging plate for high energy electron spectrometer}.
\bjtitle{Rev. Sci. Instrum.}
\bvolume{76}(\bissue{1}),
\bfpage{013507}
(\byear{2005}).
\doiurl{10.1063/1.1824371}
\end{barticle}
\endbibitem

\bibitem{bib34}
\begin{barticle}
\bauthor{\bsnm{Jeon}, \binits{J.H.}},
\bauthor{\bsnm{Nakajima}, \binits{K.}},
\bauthor{\bsnm{Kim}, \binits{H.T.}},
\bauthor{\bsnm{Rhee}, \binits{Y.J.}},
\bauthor{\bsnm{Pathak}, \binits{V.B.}},
\bauthor{\bsnm{Cho}, \binits{M.H.}},
\bauthor{\bsnm{Shin}, \binits{J.H.}},
\bauthor{\bsnm{Yoo}, \binits{B.J.}},
\bauthor{\bsnm{Jo}, \binits{S.H.}},
\bauthor{\bsnm{Shin}, \binits{K.W.}},
\bauthor{\bsnm{Hojbota}, \binits{C.}},
\bauthor{\bsnm{Bae}, \binits{L.J.}},
\bauthor{\bsnm{Jung}, \binits{J.}},
\bauthor{\bsnm{Cho}, \binits{M.S.}},
\bauthor{\bsnm{Sung}, \binits{J.H.}},
\bauthor{\bsnm{Lee}, \binits{S.K.}},
\bauthor{\bsnm{Cho}, \binits{B.I.}},
\bauthor{\bsnm{Choi}, \binits{I.W.}},
\bauthor{\bsnm{Nam}, \binits{C.H.}}:
\batitle{Measurement of angularly dependent spectra of betatron gamma-rays from
  a laser plasma accelerator with quadrant-sectored range filters}.
\bjtitle{Phys. Plasmas}
\bvolume{23}(\bissue{7}),
\bfpage{073105}
(\byear{2016}).
\doiurl{10.1063/1.4956447}
\end{barticle}
\endbibitem

\bibitem{bib35}
\begin{barticle}
\bauthor{\bsnm{D\"{o}pp}, \binits{A.}},
\bauthor{\bsnm{Hehn}, \binits{L.}},
\bauthor{\bsnm{G\"{o}tzfried}, \binits{J.}},
\bauthor{\bsnm{Wenz}, \binits{J.}},
\bauthor{\bsnm{Gilljohann}, \binits{M.}},
\bauthor{\bsnm{Ding}, \binits{H.}},
\bauthor{\bsnm{Schindler}, \binits{S.}},
\bauthor{\bsnm{Pfeiffer}, \binits{F.}},
\bauthor{\bsnm{Karsch}, \binits{S.}}:
\batitle{Quick x-ray microtomography using a laser-driven betatron source}.
\bjtitle{Optica}
\bvolume{5}(\bissue{2}),
\bfpage{199}
(\byear{2018}).
\doiurl{10.1364/OPTICA.5.000199}
\end{barticle}
\endbibitem

\bibitem{bib36}
\begin{barticle}
\bauthor{\bsnm{King}, \binits{P.M.}},
\bauthor{\bsnm{Lemos}, \binits{N.}},
\bauthor{\bsnm{Shaw}, \binits{J.L.}},
\bauthor{\bsnm{Milder}, \binits{A.L.}},
\bauthor{\bsnm{Marsh}, \binits{K.A.}},
\bauthor{\bsnm{Pak}, \binits{A.}},
\bauthor{\bsnm{Hegelich}, \binits{B.M.}},
\bauthor{\bsnm{Michel}, \binits{P.}},
\bauthor{\bsnm{Moody}, \binits{J.}},
\bauthor{\bsnm{Joshi}, \binits{C.}},
\bauthor{\bsnm{Albert}, \binits{F.}}:
\batitle{X-ray analysis methods for sources from self-modulated laser wakefield
  acceleration driven by picosecond lasers}.
\bjtitle{Rev. Sci. Instrum.}
\bvolume{90}(\bissue{3}),
\bfpage{033503}
(\byear{2019}).
\doiurl{10.1063/1.5082965}
\end{barticle}
\endbibitem

\bibitem{bib37}
\begin{barticle}
\bauthor{\bsnm{Jeon}, \binits{J.H.}},
\bauthor{\bsnm{Nakajima}, \binits{K.}},
\bauthor{\bsnm{Kim}, \binits{H.T.}},
\bauthor{\bsnm{Rhee}, \binits{Y.J.}},
\bauthor{\bsnm{Pathak}, \binits{V.B.}},
\bauthor{\bsnm{Cho}, \binits{M.H.}},
\bauthor{\bsnm{Shin}, \binits{J.H.}},
\bauthor{\bsnm{Yoo}, \binits{B.J.}},
\bauthor{\bsnm{Hojbota}, \binits{C.}},
\bauthor{\bsnm{Jo}, \binits{S.H.}},
\bauthor{\bsnm{Shin}, \binits{K.W.}},
\bauthor{\bsnm{Sung}, \binits{J.H.}},
\bauthor{\bsnm{Lee}, \binits{S.K.}},
\bauthor{\bsnm{Cho}, \binits{B.I.}},
\bauthor{\bsnm{Choi}, \binits{I.W.}},
\bauthor{\bsnm{Nam}, \binits{C.H.}}:
\batitle{A broadband gamma-ray spectrometry using novel unfolding algorithms
  for characterization of laser wakefield-generated betatron radiation}.
\bjtitle{Rev. Sci. Instrum.}
\bvolume{86}(\bissue{12}),
\bfpage{123116}
(\byear{2015}).
\doiurl{10.1063/1.4939014}
\end{barticle}
\endbibitem

\bibitem{bib38}
\begin{barticle}
\bauthor{\bsnm{Hannasch}, \binits{A.}},
\bauthor{\bsnm{Laso~Garcia}, \binits{A.}},
\bauthor{\bsnm{LaBerge}, \binits{M.}},
\bauthor{\bsnm{Zgadzaj}, \binits{R.}},
\bauthor{\bsnm{K{\"o}hler}, \binits{A.}},
\bauthor{\bsnm{Couperus~Cabada{\u{g}}}, \binits{J.P.}},
\bauthor{\bsnm{Zarini}, \binits{O.}},
\bauthor{\bsnm{Kurz}, \binits{T.}},
\bauthor{\bsnm{Ferrari}, \binits{A.}},
\bauthor{\bsnm{Molodtsova}, \binits{M.}},
\bauthor{\bsnm{Naumann}, \binits{L.}},
\bauthor{\bsnm{Cowan}, \binits{T.E.}},
\bauthor{\bsnm{Schramm}, \binits{U.}},
\bauthor{\bsnm{Irman}, \binits{A.}},
\bauthor{\bsnm{Downer}, \binits{M.C.}}:
\batitle{Compact spectroscopy of ke{V} to {M}e{V} {X}-rays from a laser
  wakefield accelerator}.
\bjtitle{Sci. Rep.}
\bvolume{11}(\bissue{1}),
\bfpage{14368}
(\byear{2021}).
\doiurl{10.1038/s41598-021-93689-5}
\end{barticle}
\endbibitem

\bibitem{bib39}
\begin{barticle}
\bauthor{\bsnm{Albert}, \binits{F.}},
\bauthor{\bsnm{Pollock}, \binits{B.B.}},
\bauthor{\bsnm{Shaw}, \binits{J.L.}},
\bauthor{\bsnm{Marsh}, \binits{K.A.}},
\bauthor{\bsnm{Ralph}, \binits{J.E.}},
\bauthor{\bsnm{Chen}, \binits{Y.-H.}},
\bauthor{\bsnm{Alessi}, \binits{D.}},
\bauthor{\bsnm{Pak}, \binits{A.}},
\bauthor{\bsnm{Clayton}, \binits{C.E.}},
\bauthor{\bsnm{Glenzer}, \binits{S.H.}},
\bauthor{\bsnm{Joshi}, \binits{C.}}:
\batitle{Angular Dependence of Betatron X-Ray Spectra from a Laser-Wakefield
  Accelerator"}.
\bjtitle{Phys. Rev. Lett.}
\bvolume{111},
\bfpage{235004}
(\byear{2013}).
\doiurl{10.1103/PhysRevLett.111.235004}
\end{barticle}
\endbibitem

\bibitem{bib40}
\begin{barticle}
\bauthor{\bsnm{Agostinelli}, \binits{S.}},
\bauthor{\bsnm{Allison}, \binits{J.}},
\bauthor{\bsnm{Amako}, \binits{K.}},
\bauthor{\bsnm{Apostolakis}, \binits{J.}},
\bauthor{\bsnm{Araujo}, \binits{H.}},
\bauthor{\bsnm{Arce}, \binits{P.}},
\bauthor{\bsnm{Asai}, \binits{M.}},
\bauthor{\bsnm{Axen}, \binits{D.}},
\bauthor{\bsnm{Banerjee}, \binits{S.}},
\bauthor{\bsnm{Barrand}, \binits{G.}},
\bauthor{\bsnm{Behner}, \binits{F.}},
\bauthor{\bsnm{Bellagamba}, \binits{L.}},
\bauthor{\bsnm{Boudreau}, \binits{J.}},
\bauthor{\bsnm{Broglia}, \binits{L.}},
\bauthor{\bsnm{Brunengo}, \binits{A.}},
\bauthor{\bsnm{Burkhardt}, \binits{H.}},
\bauthor{\bsnm{Chauvie}, \binits{S.}},
\bauthor{\bsnm{Chuma}, \binits{J.}},
\bauthor{\bsnm{Chytracek}, \binits{R.}},
\bauthor{\bsnm{Cooperman}, \binits{G.}},
\bauthor{\bsnm{Cosmo}, \binits{G.}},
\bauthor{\bsnm{Degtyarenko}, \binits{P.}},
\bauthor{\bsnm{Dell'Acqua}, \binits{A.}},
\bauthor{\bsnm{Depaola}, \binits{G.}},
\bauthor{\bsnm{Dietrich}, \binits{D.}},
\bauthor{\bsnm{Enami}, \binits{R.}},
\bauthor{\bsnm{Feliciello}, \binits{A.}},
\bauthor{\bsnm{Ferguson}, \binits{C.}},
\bauthor{\bsnm{Fesefeldt}, \binits{H.}},
\bauthor{\bsnm{Folger}, \binits{G.}},
\bauthor{\bsnm{Foppiano}, \binits{F.}},
\bauthor{\bsnm{Forti}, \binits{A.}},
\bauthor{\bsnm{Garelli}, \binits{S.}},
\bauthor{\bsnm{Giani}, \binits{S.}},
\bauthor{\bsnm{Giannitrapani}, \binits{R.}},
\bauthor{\bsnm{Gibin}, \binits{D.}},
\bauthor{\bsnm{{Gómez Cadenas}}, \binits{J.J.}},
\bauthor{\bsnm{González}, \binits{I.}},
\bauthor{\bsnm{{Gracia Abril}}, \binits{G.}},
\bauthor{\bsnm{Greeniaus}, \binits{G.}},
\bauthor{\bsnm{Greiner}, \binits{W.}},
\bauthor{\bsnm{Grichine}, \binits{V.}},
\bauthor{\bsnm{Grossheim}, \binits{A.}},
\bauthor{\bsnm{Guatelli}, \binits{S.}},
\bauthor{\bsnm{Gumplinger}, \binits{P.}},
\bauthor{\bsnm{Hamatsu}, \binits{R.}},
\bauthor{\bsnm{Hashimoto}, \binits{K.}},
\bauthor{\bsnm{Hasui}, \binits{H.}},
\bauthor{\bsnm{Heikkinen}, \binits{A.}},
\bauthor{\bsnm{Howard}, \binits{A.}},
\bauthor{\bsnm{Ivanchenko}, \binits{V.}},
\bauthor{\bsnm{Johnson}, \binits{A.}},
\bauthor{\bsnm{Jones}, \binits{F.W.}},
\bauthor{\bsnm{Kallenbach}, \binits{J.}},
\bauthor{\bsnm{Kanaya}, \binits{N.}},
\bauthor{\bsnm{Kawabata}, \binits{M.}},
\bauthor{\bsnm{Kawabata}, \binits{Y.}},
\bauthor{\bsnm{Kawaguti}, \binits{M.}},
\bauthor{\bsnm{Kelner}, \binits{S.}},
\bauthor{\bsnm{Kent}, \binits{P.}},
\bauthor{\bsnm{Kimura}, \binits{A.}},
\bauthor{\bsnm{Kodama}, \binits{T.}},
\bauthor{\bsnm{Kokoulin}, \binits{R.}},
\bauthor{\bsnm{Kossov}, \binits{M.}},
\bauthor{\bsnm{Kurashige}, \binits{H.}},
\bauthor{\bsnm{Lamanna}, \binits{E.}},
\bauthor{\bsnm{Lampén}, \binits{T.}},
\bauthor{\bsnm{Lara}, \binits{V.}},
\bauthor{\bsnm{Lefebure}, \binits{V.}},
\bauthor{\bsnm{Lei}, \binits{F.}},
\bauthor{\bsnm{Liendl}, \binits{M.}},
\bauthor{\bsnm{Lockman}, \binits{W.}},
\bauthor{\bsnm{Longo}, \binits{F.}},
\bauthor{\bsnm{Magni}, \binits{S.}},
\bauthor{\bsnm{Maire}, \binits{M.}},
\bauthor{\bsnm{Medernach}, \binits{E.}},
\bauthor{\bsnm{Minamimoto}, \binits{K.}},
\bauthor{\bsnm{{Mora de Freitas}}, \binits{P.}},
\bauthor{\bsnm{Morita}, \binits{Y.}},
\bauthor{\bsnm{Murakami}, \binits{K.}},
\bauthor{\bsnm{Nagamatu}, \binits{M.}},
\bauthor{\bsnm{Nartallo}, \binits{R.}},
\bauthor{\bsnm{Nieminen}, \binits{P.}},
\bauthor{\bsnm{Nishimura}, \binits{T.}},
\bauthor{\bsnm{Ohtsubo}, \binits{K.}},
\bauthor{\bsnm{Okamura}, \binits{M.}},
\bauthor{\bsnm{O'Neale}, \binits{S.}},
\bauthor{\bsnm{Oohata}, \binits{Y.}},
\bauthor{\bsnm{Paech}, \binits{K.}},
\bauthor{\bsnm{Perl}, \binits{J.}},
\bauthor{\bsnm{Pfeiffer}, \binits{A.}},
\bauthor{\bsnm{Pia}, \binits{M.G.}},
\bauthor{\bsnm{Ranjard}, \binits{F.}},
\bauthor{\bsnm{Rybin}, \binits{A.}},
\bauthor{\bsnm{Sadilov}, \binits{S.}},
\bauthor{\bsnm{{Di Salvo}}, \binits{E.}},
\bauthor{\bsnm{Santin}, \binits{G.}},
\bauthor{\bsnm{Sasaki}, \binits{T.}},
\bauthor{\bsnm{Savvas}, \binits{N.}},
\bauthor{\bsnm{Sawada}, \binits{Y.}},
\bauthor{\bsnm{Scherer}, \binits{S.}},
\bauthor{\bsnm{Sei}, \binits{S.}},
\bauthor{\bsnm{Sirotenko}, \binits{V.}},
\bauthor{\bsnm{Smith}, \binits{D.}},
\bauthor{\bsnm{Starkov}, \binits{N.}},
\bauthor{\bsnm{Stoecker}, \binits{H.}},
\bauthor{\bsnm{Sulkimo}, \binits{J.}},
\bauthor{\bsnm{Takahata}, \binits{M.}},
\bauthor{\bsnm{Tanaka}, \binits{S.}},
\bauthor{\bsnm{Tcherniaev}, \binits{E.}},
\bauthor{\bsnm{{Safai Tehrani}}, \binits{E.}},
\bauthor{\bsnm{Tropeano}, \binits{M.}},
\bauthor{\bsnm{Truscott}, \binits{P.}},
\bauthor{\bsnm{Uno}, \binits{H.}},
\bauthor{\bsnm{Urban}, \binits{L.}},
\bauthor{\bsnm{Urban}, \binits{P.}},
\bauthor{\bsnm{Verderi}, \binits{M.}},
\bauthor{\bsnm{Walkden}, \binits{A.}},
\bauthor{\bsnm{Wander}, \binits{W.}},
\bauthor{\bsnm{Weber}, \binits{H.}},
\bauthor{\bsnm{Wellisch}, \binits{J.P.}},
\bauthor{\bsnm{Wenaus}, \binits{T.}},
\bauthor{\bsnm{Williams}, \binits{D.C.}},
\bauthor{\bsnm{Wright}, \binits{D.}},
\bauthor{\bsnm{Yamada}, \binits{T.}},
\bauthor{\bsnm{Yoshida}, \binits{H.}},
\bauthor{\bsnm{Zschiesche}, \binits{D.}}:
\batitle{Geant4—a simulation toolkit}.
\bjtitle{Nucl. Instrum. Methods Phys. Res. Sect. A}
\bvolume{506}(\bissue{3}),
\bfpage{250}
(\byear{2003}).
\doiurl{10.1016/S0168-9002(03)01368-8}
\end{barticle}
\endbibitem

\bibitem{bib41}
\begin{barticle}
\bauthor{\bsnm{King}, \binits{P.M.}},
\bauthor{\bsnm{Rusby}, \binits{D.}},
\bauthor{\bsnm{Hannasch}, \binits{A.}},
\bauthor{\bsnm{Lemos}, \binits{N.}},
\bauthor{\bsnm{Tiwari}, \binits{G.}},
\bauthor{\bsnm{Pak}, \binits{A.}},
\bauthor{\bsnm{Kerr}, \binits{S.}},
\bauthor{\bsnm{Cochran}, \binits{G.}},
\bauthor{\bsnm{Pagano}, \binits{I.}},
\bauthor{\bsnm{Williams}, \binits{G.J.}},
\bauthor{\bsnm{Khan}, \binits{S.F.}},
\bauthor{\bsnm{Aufderheide}, \binits{M.}},
\bauthor{\bsnm{Kemp}, \binits{A.}},
\bauthor{\bsnm{Wilks}, \binits{S.}},
\bauthor{\bsnm{Macphee}, \binits{A.}},
\bauthor{\bsnm{Albert}, \binits{F.}},
\bauthor{\bsnm{Hegelich}, \binits{B.M.}},
\bauthor{\bsnm{Downer}, \binits{M.}},
\bauthor{\bsnm{Manuel}, \binits{M.}},
\bauthor{\bsnm{Gavin}, \binits{Z.}},
\bauthor{\bsnm{Haid}, \binits{A.}},
\bauthor{\bsnm{Mackinnon}, \binits{A.}}:
\batitle{Enhancement of high energy X-ray radiography using compound parabolic
  concentrator targets}.
\bjtitle{High Energy Density Phys.}
\bvolume{42},
\bfpage{100978}
(\byear{2022}).
\doiurl{10.1016/j.hedp.2022.100978}
\end{barticle}
\endbibitem

\bibitem{bib42}
\begin{barticle}
\bauthor{\bsnm{Guin}, \binits{U.}},
\bauthor{\bsnm{Huang}, \binits{K.}},
\bauthor{\bsnm{DiMase}, \binits{D.}},
\bauthor{\bsnm{Carulli}, \binits{J.M.}},
\bauthor{\bsnm{Tehranipoor}, \binits{M.}},
\bauthor{\bsnm{Makris}, \binits{Y.}}:
\batitle{Counterfeit Integrated Circuits:A Rising Threat in the Global
  Semiconductor Supply Chain}.
\bjtitle{Proceedings of the IEEE}
\bvolume{102}(\bissue{8}),
\bfpage{1207}
(\byear{2014}).
\doiurl{10.1109/JPROC.2014.2332291}
\end{barticle}
\endbibitem

\bibitem{bib43}
\begin{botherref}
Datasheet, Hamamatsu 90 keV microfocus x-ray source L9421-02.
\url{https://www.hamamatsu.com/content/dam/hamamatsu-photonics/sites/documents/99_SALES_LIBRARY/etd/L9421-02_TXPR1011E.pdf}
(Acces date: 2022.11.14)
\end{botherref}
\endbibitem

\bibitem{bib44}
\begin{bchapter}
\bauthor{\bsnm{Ahi}, \binits{K.}},
\bauthor{\bsnm{Asadizanjani}, \binits{N.}},
\bauthor{\bsnm{Shahbazmohamadi}, \binits{S.}},
\bauthor{\bsnm{Tehranipoor}, \binits{M.}},
\bauthor{\bsnm{Anwar}, \binits{M.}}:
\bctitle{{Terahertz characterization of electronic components and comparison of
  terahertz imaging with x-ray imaging techniques}}.
In: \beditor{\bsnm{Anwar}, \binits{M.F.}},
\beditor{\bsnm{Crowe}, \binits{T.W.}},
\beditor{\bsnm{Manzur}, \binits{T.}} (eds.)
\bbtitle{Terahertz Physics, Devices, and Systems IX: Advanced Applications in
  Industry and Defense},
vol. \bseriesno{9483},
p. \bfpage{94830}.
\bpublisher{SPIE}, \blocation{ }
(\byear{2015}).
\bcomment{International Society for Optics and Photonics}.
\burl{https://doi.org/10.1117/12.2183128}
\end{bchapter}
\endbibitem

\bibitem{bib45}
\begin{barticle}
\bauthor{\bsnm{Soto}, \binits{L.}},
\bauthor{\bsnm{Pavez}, \binits{C.}},
\bauthor{\bsnm{Moreno}, \binits{J.}},
\bauthor{\bsnm{Cárdenas}, \binits{M.}},
\bauthor{\bsnm{Tarifeño}, \binits{A.}},
\bauthor{\bsnm{Silva}, \binits{P.}},
\bauthor{\bsnm{Zambra}, \binits{M.}},
\bauthor{\bsnm{Huerta}, \binits{L.}},
\bauthor{\bsnm{Tenreiro}, \binits{C.}},
\bauthor{\bsnm{Giordano}, \binits{J.L.}},
\bauthor{\bsnm{Lagos}, \binits{M.}},
\bauthor{\bsnm{Retamal}, \binits{C.}},
\bauthor{\bsnm{Escobar}, \binits{R.}},
\bauthor{\bsnm{Ramos}, \binits{J.}},
\bauthor{\bsnm{Altamirano}, \binits{L.}}:
\batitle{Dense transient pinches and pulsed power technology: research and
  applications using medium and small devices}.
\bjtitle{Phys. Scripta}
\bvolume{2008}(\bissue{T131}),
\bfpage{014031}
(\byear{2008}).
\doiurl{10.1088/0031-8949/2008/T131/014031}
\end{barticle}
\endbibitem

\bibitem{bib46}
\begin{barticle}
\bauthor{\bsnm{Raspa}, \binits{V.}},
\bauthor{\bsnm{Moreno}, \binits{C.}},
\bauthor{\bsnm{Sigaut}, \binits{L.}},
\bauthor{\bsnm{Clausse}, \binits{A.}}:
\batitle{Effective hard x-ray spectrum of a tabletop Mather-type plasma focus
  optimized for flash radiography of metallic objects}.
\bjtitle{J. Applied Phys.}
\bvolume{102}(\bissue{12}),
\bfpage{123303}
(\byear{2007}).
\doiurl{10.1063/1.2822449}
\end{barticle}
\endbibitem

\end{thebibliography}


\end{document}